%% file: master_CoRR.tex
\documentclass[12pt]{book}

\usepackage{amsfonts}
\usepackage{amssymb}
\usepackage{amsthm}
\usepackage{simplethesis2}
\usepackage{epsfig}

\DeclareSymbolFontAlphabet{\mathbb}{AMSb}

\long\def\ignore#1{}

\newcommand{\ra}{\rightarrow}
\newcommand{\ie}{\emph{i.e.}}

\newcommand{\lambdase}{\lambda s_e}
\newcommand{\lambdasi}{\lambda\sigma}
\newcommand{\lambdaup}{\lambda\upsilon}
\newcommand{\lambdax}[1]{\lambda #1 \,}
\newcommand{\lambdadb}{\lambda \,}
\newcommand{\allx}{\forall}
\newcommand{\andxy}[2]{#1 \wedge #2}

\newcommand{\app}{{\ }}
\newcommand{\dum}[1]{@ #1}
\newcommand{\rnf}[1]{{\vert #1 \vert}}
\newcommand{\lenv}{{\lbrack\!\lbrack}}
\newcommand{\renv}{{\rbrack\!\rbrack}}
\newcommand{\env}[1]{{\lenv #1 \renv}}

\newtheorem{defn}{Definition}[subsection]
\newtheorem{theorem}[defn]{Theorem}
\newtheorem{corollary}[defn]{Corollary}
\newtheorem{lemma}[defn]{Lemma}
\newtheorem{theorem1}{Theorem}[section]
\newtheorem{theorem2}{Theorem}[chapter]

\begin{document}

\setcounter{secnumdepth}{5}

\title{Reduction Strategies in Lambda Term Normalization and their
Effects on Heap Usage\thanks{This article is a modified version of
the master's thesis of Xiaochu Qi}}

\author{Xiaochu Qi \\ Computer Science and Engineering \\ University of Minnesota\\
                      4-192 EE/CS Building\\ 200 Union Street S.E. \\ Minneapolis, MN 55455\\
                      xqi@cs.umn.edu}

\maketitle

\pagenumbering{roman} \abstract{\input{abs}}
\acknowledgments{\input{ack}}
\setcounter{tocdepth}{2}

\tableofcontents \clearpage \listoffigures \pagenumbering{arabic}

\textpages

\singlespace

\input{intro}
\input{background}
\input{implicit}

\input{naive}
\input{combined}

\input{comparison1}
\input{conclusion}
\bibliographystyle{plain}
\bibliography{master}

\end{document}

%% file: abs.tex
Higher-order representations of objects such as programs, proofs,
formulas and types have become important to many symbolic
computation tasks. Systems that support such representations
usually depend on the implementation of an intensional view of the
terms of some variant of the typed $\lambda$-calculus. Various
notations have been proposed for $\lambda$-terms to explicitly
treat substitutions as basis for realizing such implementations.
There are, however, several choices in the actual reduction
strategies. The most common strategy utilizes such notations only
implicitly via an incremental use of environments. This approach
does not allow the smaller substitution steps to be intermingled
with other operations of interest on $\lambda$-terms. However, a
naive strategy explicitly using such notations can also be costly:
each use of the substitution propagation rules causes the creation
of a new structure on the heap that is often discarded in the
immediately following step. There is thus a tradeoff between these
two approaches. This thesis describes the actual realization of the 
two approaches, discusses their tradeoffs based on this and, finally, 
offers an amalgamated approach that utilizes recursion in rewrite rule
application but also suspends substitution operations where
necessary.

%% file: ack.tex
I take this opportunity to express my gratitude to my advisor Dr.
Gopalan Nadathur for his guidance, support, and valuable instruction
throughout this entire project. I am grateful to Dr. Chuck C. Liang
for his help and interest in this project. Thanks are also due to Dr. 
Eric Van Wyk and Dr. William Messing for their interest in this project.  

I would also like to thank Natalie Linnell, Lijesh Krishnan, and Jimin Gao
for their comments on drafts and this thesis.

Work on this thesis has been partially supported by NSF Grant 
CCR-0096322, a Grant-in-Aid of Research from the University of Minnesota 
and by funds provided by the Institute of Technology and the Department 
of Computer Science and Engineering at the University of Minnesota.

%% file: intro.tex
\chapter{Introduction}\label{chp:intro}

This thesis is concerned with the treatment of $\lambda$-terms in a
situation where they are used as a means for representing
syntactic objects whose structures involve binding. Such a use
occurs in a variety of metaprogramming and symbolic computation
tasks, such as proof assistants
\cite{deBruijn80,Cons86,Dowek93tr,Paulson94}, logical frameworks
\cite{CH88,HHP93} and metalanguages \cite{NM88,Pfenning94cade}.
The usefulness of $\lambda$-terms in representing higher-order
syntactic objects, $\ie$ objects involving binding, lies in the
following two aspects. First, the binding notions can be encoded
by $\lambda$-abstractions explicitly. For example, in a theorem
proving context, the first-order logic formula $\allx{x}\app
P(x)$, where $P(x)$ represents an arbitrary formula in which $x$
perhaps appears free, can be encoded by the $\lambda$-term
$(all\app(\lambdax{x}\app\overline{P(x)}))$, where
$\overline{P(x)}$ denotes, recursively, the representation of
$P(x)$, and the constructor $all$ encodes the universal quantifier.
The binding scope of the universal quantifier over $x$ in the
above formula is explicitly represented by the
$\lambda$-abstraction. Second, the important substitution
computation on such objects is captured by the attendant
$\beta$-reduction operation on $\lambda$-terms. For example, in
the formula above, the instantiation of the quantifier over $x$
with a term denoted by $t$ can be simply represented as
$(\lambdax{x}\app\overline{P(x)})\app t$. The displayed
$\lambda$-term can be rewritten to a form in which the occurrences
of $x$ in $\overline{P(x)}$ have been replaced by $t$ using the
$\beta$-reduction operation, and therefore represents the result
of substituting $t$ properly for $x$ when we assume a notion of
equality on $\lambda$-terms based on this operation.

Since $\lambda$-terms are used as data structures to represent
syntactic objects, these terms may often have to be compared in
the course of symbolic computation over such objects. To perform
comparisons properly, the equality relation between
$\lambda$-terms must incorporate variable renaming, $\ie$
$\alpha$-conversion. For example, the two $\lambda$-terms
$(all\app(\lambdax{x}\app\overline{P(x)}))$ and
$(all\app(\lambdax{y}\app\overline{P(y)}))$ which represent
formulae $\allx{x}\app P(x)$ and $\allx{y}\app P(y)$ respectively,
should be recognized as being equal. Further, as we have noted
already, in determining equality, it is necessary to take into
account the notion of $\beta$-conversion. For instance, in a
theorem proving context, we may want to check whether the formula
generated from instantiating the quantifier over $x$ in the
formula
\begin{tabbing}[t]
\qquad\=\kill \>
$\allx{x}\app(\allx{y}\app(\andxy{P(x,y)}{Q(x,y)}))$
\end{tabbing}
is a universally quantified formula with a conjunction as its
top-level connective inside the quantifier. This computation
requires matching the $\lambda$-term
\begin{tabbing}[t]
\qquad\=\kill \> $(\lambdax{x}\app (all\app(\lambdax{y}\app
(and\app(\overline{P(x,y)},\overline{Q(x,y)})))))\app t$
\end{tabbing}
with a term of the form $(all\app(\lambdax{z}\app
(and\app(R,T))))$, where $R$ and $T$ are schema variables that may
be instantiated in the course of the matching; we note that we are
using the constructor $and$ here to encode the logical connective
$\land$. An important point to note here is that in order for the
matching to be performed, the $\beta$-redex in the first term has
to be contracted and the top-level structure of the resulting
formula has to be exposed. The latter actually means that it is
necessary to propagate a substitution into a context embedded
within an abstraction. This is an aspect that is novel to the use
of $\lambda$-terms as representational devices. In the context of
functional programming, for instance, it is never necessary to
look inside abstraction contexts.

Considering the frequency with which it is used, the efficiency of
the implementation of the $\beta$-reduction operation has a
significant impact on the performance of the system that supports
the use of $\lambda$-terms as its data structures. Focusing on
this issue in the realization of such systems is therefore
important. A significant part of the $\beta$-reduction operation
is substitution. Traditional presentations of $\beta$-contraction,
the single step rewriting process from which the $\beta$-reduction
operation is constructed, take a rather simplistic view of
substitution. The operation is usually presented via a rule such
as
\begin{tabbing}[t]
\qquad\=\kill \>$(\lambdax{x}\app t_1)\app t_2 \ \rightarrow \
t_1[x:=t_2]$,
\end{tabbing}
\noindent where $t_1[x:=t_2]$ denotes the term obtained by
replacing the free occurrences of $x$ in $t_1$ by $t_2$, carrying
out the necessary renaming in the process to ensure that binding
scopes are properly respected. Unfortunately, from an
implementation perspective, the substitution operation is too
complex to be performed as an atomic step. In particular, this
substitution operation requires going through the structure of
$t_1$, taking care to rename the bound variables where necessary
to avoid illegal capture, and eventually replacing free
occurrences of $x$ with $t_2$. For this reason, in real
implementations, the substitution operation is often broken into
smaller steps. Each step focuses on one specific substructure of
the term. For example, consider the term $(\lambdax{x}\app
(t_1\app(\lambdax{y}\app t_2)))\app t_3$. When it is observed that
this term can be rewritten using the $\beta$-contraction
rule, the reduction process registers the substitution of
$t_3$ for $x$ in an environment. Calling this environment $e$, the
task now becomes that of propagating it over $t_1$ and
$\lambdax{y}\app t_2$. In processing the term with the
abstraction, it may become necessary to rename the bound variable
and this can be built into the rule for this case, possibly
resulting in the addition of the substitution of, say, $z$ for $y$ to
the environment $e$. Finally, in traversing $t_1$ and $t_2$, a
variable may be encountered and the result in this case would be
to possibly substitute a term based on the environment.

The use of an environment actually leads to some possible
improvements in the implementation of reduction. First, the
delaying of substitutions gives us the ability to combine
substitutions generated by different $\beta$-contractions into the
\emph{same} environment so that they can all be performed in one
traversal over the involved term. For example, consider the term
$(\lambdax{x}\app\lambdax{y}\app t_1)\app t_2\app t_3$. There are
two redices that have to be contracted in normalizing this term.
If the naive approach to $\beta$-contraction is used, it would be
necessary to walk through the structure of $t_1$ twice in
effecting the necessary substitutions. With the delaying of
substitutions, the two substitutions $[x:=t_2]$ and $[y:=t_3]$ can
be combined into one environment and performed in the same
traversal over $t_1$. The delaying of substitutions also gives us
more opportunities to avoid unnecessary term traversals. For
example, in the term considered above, the variable $y$ that is to
be substituted for by $t_3$ can only occur inside $t_1$. Under
the naive approach to $\beta$-contraction, we would first
substitute $t_2$ for all the free occurrences of $x$ in $t_1$.
Then the traversal to substitute $t_3$ for $y$ would also examine
these (new) occurrences of $t_2$ to see if there are any free
occurrences of $y$ in them to replace. However, there are \emph{no}
such $y$'s in $t_2$. With the delayed performance of substitution,
such substitution traversals can be recognized and avoided. This
kind of approach to substitution and reduction has, in fact been
central to the implementation of functional programming languages.
In recent years, it has also been given a formal basis by the work
on explicit substitution notations (e.g. see
\cite{ACCL91,KR97,NW98tcs}) that incorporate the possibility of
encoding suspended substitutions in terms. These notations also
make it possible to extend this approach to reduction even to
situations where we need to look inside abstractions.

The above discussion leads naturally to an implementation of
$\beta$-reduction that is environment based. In the simplest 
form, such a procedure would be guided by an explicit substitution
notation but would use suspended substitutions only implicitly,
$\ie$ it would not explicitly create terms with suspended
substitutions as their (sub)structures, but, instead, it would
record those suspended substitutions via local variables and
parameters of the reduction procedure. Thus the terms eventually
produced by such a procedure would not contain subparts encoding
suspended substitutions. More specifically, when the non-reducible
head of a term is found, the procedure would need to actually
carry out the suspended substitutions on the remaining part of the
term structure. The consequence of this is that the ability to
delay and combine substitutions is limited to the extent of a
single invocation of the reduction procedure, which means the
opportunities for sharing the substitution walks that are caused
by contracting redices generated dynamically by other kinds of
computation steps are missed. For example, consider a formula in the 
first-order logic that is represented by
$(all\app (\lambdax{x}\app\ldots (all\app(\lambdax{y}\app P))
\ldots))$, where $P$ is an arbitrary $\lambda$-term. Suppose that
we now carry out a computation over this formula that involves
recognizing and instantiating all its universal quantifiers. This
computation would first recognize the outermost quantifier and
instantiate it by generating the term
$(\lambdax{x}\app\ldots(all\app(\lambdax{y}\app P))\ldots)\app t$.
Now the outer redex would be contracted, generating a substitution
computation involving the variable $x$. At a later stage, the
inner quantifier will be noticed, generating another substitution
computation, this time involving the variable $y$. Thus, it is
necessary in principle to substitute for two different variables
in the structure represented by $P$. If we do not have the ability
to delay substitutions beyond the extent of one invocation of the
reduction procedure, each of these substitutions requires a
separate walk over the structure of $P$. This leads to overhead
in both processing time and in the creation of term structures.

To overcome this kind of overhead, it becomes meaningful to
consider a reduction procedure that uses the ability to suspend
substitutions explicitly by sometimes returning terms that have
(sub)structures encoding other terms with substitutions yet to be
performed on them. Now, explicit substitution notations are
usually presented via rules that generate and propagate
substitutions. For example, the $\beta$-contraction operation may
be expressed via a rule of the form
\begin{tabbing}[t]
\qquad\=\kill \>$(\lambdax{x}\app t_1)\app t_2 \ \rightarrow \
[t_1, (x,t_2)::nil]$,
\end{tabbing}
where an expression of the form $[t_1,e]$ represents the term
$t_1$ with substitutions contained in the environment $e$ to be
performed on it, and environments are represented as lists of
bindings. Similarly, we may have a rule of the form
\begin{tabbing}[t]
\qquad\=\kill \>$[(t_1\app t_2), e] \ \rightarrow \ [t_1,e]\app
[t_2,e]$
\end{tabbing}
that realizes the propagation of substitutions over applications.
Assuming such a presentation, the simplest way of realizing the
kind of reduction procedure we desire would be to use these kinds
of rules directly, explicitly creating structures corresponding to
the righthand sides when matching those corresponding to the
lefthand sides. In the end, when the top-level structure of the
term has been exposed to the extent desired, $\ie$ the
non-reducible head of the term is found, the rewriting process is
stopped. Notice that in this process some subterms may be left in
the form of $[t,e]$ to be further evaluated as desired at some
later stage of computation.

 The second strategy that we have described  clearly solves the
problems noted for the first strategy but it also has problems of its own.
In particular, it has the potential drawback for creating
\emph{too} many new structures. This would happen if, for
instance, the righthand sides of rules that we create become the
lefthand sides of other rules and have to be rewritten
immediately. Now, new terms that are created have obviously to be
allocated in dynamic space, $\ie$ in the heap. If a lot of space
is unnecessarily allocated in the heap, it will become necessary
in an industrial-strength system to reclaim such space using a
garbage collector. The running time of the garbage collector
becomes a factor in the overall performance of the computational
system, and therefore we would like to reduce it as much as
possible.

The organization of the first reduction procedure based on
environment suggests a way to avoid such a potentially profligate
use of heap space. Rather than creating new terms with embedded
substitutions immediately, the suspended substitutions may be
represented implicitly by the parameters and local variables of
the reduction procedure. However when the non-reducible head of a
term is found, rather than performing the substitutions on the
remaining term structures right away, new structures that maintain
such substitutions in suspended forms can be created explicitly.
This approach combines the implicit and explicit treatments of
substitutions and can accrue the benefits of both.

In earlier paragraphs, we have outlined three different
approaches to realizing reduction and have discussed their
potential drawbacks and advantages. In this thesis we lend
concreteness to these informal discussions. Our particular
contributions are twofold:
\begin{enumerate}
\item Using a specific explicit substitution notation~\cite{nadathur99finegrained}, we
develop the three strategies into reduction procedures that can be
embedded in actual systems. The development of the third strategy
is new to our work and that of the first extends usual environment
based procedures to a situation where it is important to look
within abstractions. We also include correctness proofs with our
procedures.
\item We quantify the differences between the
different strategies through experiments on ``real-life''
computations. To conduct this study, we have realized our
reduction procedures in the C language and have embedded them
within the Teyjus implementation~\cite{NM99cade} of
$\lambda$Prolog~\cite{NM88}---a language that provides
$\lambda$-terms as data structures---and have collected data from
a suite of user programs using the resulting versions of the
system. We believe this kind of a study to be unique to our work
and its encompassing project~\cite{NX03,LNX03}.
\end{enumerate}

The rest of this thesis is organized as follows. In the next
chapter we introduce the $\lambda$-calculus and describe an explicit
substitution calculus called the suspension notation
\cite{NW98tcs}; this notation has already been used in two
practical systems \cite{NM99cade,shao98:imp} and is therefore an
appealing basis for our study. Chapters~\ref{chp:implicit},
\ref{chp:naive} and \ref{chp:combined} then present reduction
procedures realizing the three different approaches of interest.
These procedures are presented using the SML language both for
simplicity of exposition and for concreteness, although the same
ideas can be deployed in realizations in any other language as
well. Chapter~\ref{chp:comparison} contains a quantitative
comparison of these approaches using, in fact, a C based
realization of each. Chapter~\ref{chp:conclusion} concludes the
thesis.

%% file: background.tex
\chapter{The $\lambda$-Calculus and Explicit Substitutions}\label{chp:background}
This chapter provides technical background needed for our later
discussions. In Section~\ref{sec:lambda-calculus}, we give an
overview of the $\lambda$-calculus. Section~\ref{sec:debruijn}
introduces the de Bruijn notation. In Section~\ref{sec:exp-sub},
we introduce the idea of explicit substitution calculi and
describe the suspension notation as a representative.
\section{An Overview of the $\lambda$-Calculus}\label{sec:lambda-calculus}
Invented by Alonzo Church in 1930's, the $\lambda$-calculus is
designed to capture the most basic aspects of functionality, $\ie$
what it is means to be a function and what it means to apply a
function to arguments under this interpretation. Using
$\lambda$-terms to represent syntactic objects naturally requires
the ability to perform comparisons between $\lambda$-terms.
The usual method is to
transform the terms to their $\beta$-normal forms first and then
to check the equality between those normal forms, which we will
discuss in detail in Section~\ref{subsec:con}. Thus the normal
forms of the terms under comparison should be guaranteed to
exist. In the context of using $\lambda$-terms to represent
syntactic objects, we are mainly interested in the use of typed
$\lambda$-calculi. In these situations, the set of terms is
restricted to only those that satisfy certain typeable constraints, 
and this restriction usually ensures the existence of
$\beta$-normal forms. However, since we are not interested in any
property associated with the types of $\lambda$-terms other than
the guarantee of the existence of $\beta$-normal forms, our
discussion in this thesis is still based on the untyped
$\lambda$-calculus for generality and simplicity.
Section~\ref{subsec:lambda-terms} introduces the terms of the
untyped $\lambda$-calculus. Section~\ref{subsec:substitutions}
presents the substitution operation. Section~\ref{subsec:con}
describes the important $\alpha$-conversion and $\beta$-conversion
operations and the equality notion of $\lambda$-terms based upon
them.

\subsection{Terms in the $\lambda$-Calculus}\label{subsec:lambda-terms}
We begin with the definition of $\lambda$-terms.
\begin{defn}\label{def:lambda-terms}
We assume in the beginning that we are given a set of {\it
constants}, a set of {\it abstractable variables} and a set of {\it
instantiatable variables}. The set of $\lambda$-terms is then the
smallest set obtained from the combination of these sets using the
following operations:
\begin{enumerate}
\item abstraction, that produces the term $(\lambdax{x}\app t)$ given
an abstractable variable $x$ and a $\lambda$-term $t$, and
\item application, that produces the term $(t_1\app t_2)$ given two
$\lambda$-terms $t_1$ and $t_2$.
\end{enumerate}
In an abstraction of the form $(\lambdax{x}\app t)$, we say that the scope
of this abstraction is $t$ and we also refer to $t$ as the body of
this abstraction. In an application of the form $(t_1\app t_2)$, we refer
to $t_1$ as the function and $t_2$ as the argument of this
application.
\end{defn}

The instantiatable variables in this definition are also referred to as 
logic variables. They differ from the abstractable variables in the sense
that they cannot be substituted by $\beta$-reductions, which we will 
discuss later, but possibly be replaced by other operations trying to 
make two $\lambda$-terms equal, $\ie$, unifications.   

For the sake of readability, we often omit parentheses surrounding
abstractions and applications when we write terms,
assuming that these can be
inserted using the following conventions: applications associate
to the left and have a higher priority than abstractions.

Intuitively, the abstraction term $(\lambdax{x}\app t)$ is
intended to represent the function that when given $x$ returns
$t$. In this sense, $x$ is the formal argument and $t$ is the body
of this function. For example, using $+$ as an infix operator, the
abstraction $(\lambdax{x}\app x+1)$ represents a function that
when given a value for $x$ returns $(x+1)$.

Now consider two $\lambda$-terms $(\lambdax{x}\app x+1)$ and
$(\lambdax{y}\app y+1)$. It can be seen that these two terms both
represent the same function: when given the same actual argument,
they return the same value, that is, the actual argument plus one.
For this reason, we want to recognize these two terms as being
equal. The equality relation that we describe for $\lambda$-terms
in Section~\ref{subsec:con} actually encompasses this idea.

The intuitive meaning of an application term is the application of
a function to actual arguments. For example, the term
$((\lambdax{x}\app x+1)\app 2)$ represents the application of the
function $(\lambdax{x}\app x+1)$ to $2$. Naturally, this
application term is equal to $2+1$. For this reason, we expect to
generally recognize the term representing the application of a
function to a value, and the term representing the result of
evaluating this application, to be equal. The notion of equality we
formalize in Section~\ref{subsec:con} also encompasses this idea.

\begin{defn}
Term $t$ is said to be an {\it immediate subterm} of $s$ when $s$
is in the form of $(\lambdax{x}\app t)$, $(t \app t_1)$ or $(t_1
\app t)$. A term $t$ is a {\it subterm of} $s$ if it is $s$ or,
recursively, a subterm of an immediate subterm of $s$.
\end{defn}

\begin{defn}
Let $t$ be a $\lambda$-term that has a subterm $(\lambdax{x}\app
t_1)$. All the occurrences of $x$ in $(\lambdax{x}\app t_1)$ are said
to be {\it bound} in $t$, and $x$ is called a {\it bound variable}
of the subterm $(\lambdax{x}\app t_1)$. Any non-bound occurrence of
$x$ is said to be {\it free} in $t$. If $x$ has at least one free
occurrence in $t$, then it is called a {\it free variable of} $t$;
the set of all the free variables of $t$ is represented by
$FV(t)$. If $FV(t)=\emptyset$, we refer to $t$ as a closed term.
\end{defn}
According to this definition, $x$ is a bound variable of the
closed term $(\lambdax{x}\app x+1)$. For another example, consider
the term $(\lambdax{y}\app(\lambdax{x}\app x+y))$. The variable
$y$ is free in its subterm $(\lambdax{x}\app x+y)$, even though it
is bound in the top-level term.

\subsection{Substitutions}\label{subsec:substitutions}
To compare two $\lambda$-terms $(\lambdax{x}\app t)$ and
$(\lambdax{y}\app s)$, it is required to rename their formal
arguments to be the same. In particular, we need to rename the
variable $y$ in $s$ by $x$ first, then further check whether the
resulting function body is the same as the one represented by $t$.
The evaluation of function application $((\lambdax{x}\app t)\app
s)$ also requires replacing the occurrences of its formal argument
$x$ by $s$ inside $t$. In this sense, these two kinds of
operations both have substitution as a main component. However,
the substitution operation is not always so straightforward and
extra attention should be paid to perform it correctly. For
example, consider the evaluation of the $\lambda$-term
$(\lambdax{y}\app((\lambdax{x}\app\lambdax{y} \app x)\app y))$,
which requires substituting the variable $y$ for the variable $x$
in its subterm $(\lambdax{y}\app x)$. Note that the occurrence of
the variable $y$ is bound by the top level abstraction but is free
in the abstraction it will be substituted in. If we directly
replace $x$ with $y$, the evaluation result would be
$(\lambdax{y}\app (\lambdax{y}\app y))$, in which $y$ is
incorrectly bound by the inner abstraction instead of the top
level one. To preserve the correct binding relation, we need to
first rename $y$ in $(\lambdax{y}\app x)$ to a new abstractable
variable $z$, and then replace $x$ with $y$ to obtain the term
$(\lambdax{y}\app (\lambdax{z}\app y))$. To present such term
replacement operations systematically, we define the substitution
of $\lambda$-terms as the following:
\begin{defn}
For any $\lambda$-terms $t$, $s$ and abstractable variable $x$, $t[x:=s]$
represents the result of substituting $s$ for every free
occurrence of $x$ in $t$ simultaneously and is defined recursively
on the structure of $t$ as the following.
\begin{enumerate}
\item If $t$ is an abstractable variable, $t[x:=s]= \left\{ \begin{array}{ll}
                             s & \mbox{if $t=x$} \\
                             t & \mbox{if $t\neq x$;}
                             \end{array}
                    \right.$
\item If $t$ is an instantiatable variable, $t[x:=s]=t$;
\item If $t$ is a constant, $t[x:=s]=t$;
\item If $t$ is an application $(t_1\app t_2)$,
$(t_1 \app t_2)[x:=s]=(t_1[x:=s])\app (t_2[x:=s])$;
\item If $t$ is an abstraction $(\lambdax{y} \app t_1)$, \\
$(\lambdax{y}\app t_1)[x:=s]= \left\{ \begin{array}{ll}
                                  \lambdax{y}\app t_1 & \mbox{if $y=x$} \\
                                  \lambdax{y}\app (t_1[y:=s]) & \mbox{if $y \neq x$ and $y \notin
                                            FV(s)$} \\
                                  \lambdax{z}\app
                                   (t_1[y:=z][x:=s]) & \mbox{if $y\neq x$, $y \in
                                            FV(s)$ } \\
                                   & \mbox{and $z\notin FV(t_1)\cup FV(s)$.}
                                      \end{array}
                                      \right. $
\end{enumerate}
\end{defn}
Among these substitution rules the last case of rule (5) takes
care of renaming bound variables to avoid illegal capture, $\ie$
name clashes.

\subsection{Rules of $\lambda$-Conversions}\label{subsec:con}
One component of the equality relation that we want to define on
$\lambda$-terms is that of recognizing the irrelevance of bound
variable names. This is formalized through the notion of
$\alpha$-conversion defined below.
\begin{defn}
Let $t$ be a $\lambda$-term that has a subterm $(\lambdax{x}\app
s)$, and let $y$ be an abstractable variable such that $y \notin
FV(s)$. The action of replacing this subterm $(\lambdax{x}\app s)$
with $(\lambdax{y}\app (s[x:=y]))$ is called an {\it
$\alpha$-conversion}. We say $t$ {\it $\alpha$-converts to} $t'$
if and only if $t'$ has been obtained from $t$ by a finite
(perhaps empty) series of $\alpha$-conversions.
\end{defn}

Note that $\alpha$-conversion is also used implicitly in the
substitution operation.

The evaluation process of a function application represented by
term $((\lambdax{x}\app t)\app s)$ is the operation of replacing
the free occurrences of $x$ inside $t$ with $s$, which can be
denoted as $t[x:=s]$.
This process is captured by the {\it $\beta$-contraction}
operation.
\begin{defn}
Let $t$ be a $\lambda$-term that has a subterm in the form of
$((\lambdax{x}\app t_1)\app t_2)$ which is called a $\beta$-redex.
The action of replacing this subterm with $(t_1[x:=t_2])$ is
called a {\it $\beta$-contraction}. We say that $t \rhd_{\beta}
t'$ if and only if $t'$ has been obtained from $t$ by a
$\beta$-contraction. A finite (perhaps empty) series of
$\beta$-contractions is called a {\it $\beta$-reduction}.
\end{defn}

Clearly, the following property is preserved by $\beta$-contraction.
\begin{lemma}\label{free variables}
Let $t$ and $s$ be $\lambda$-terms. If $t \rhd_{\beta} s$,
then $FV(s)\subseteq FV(t)$.
\end{lemma}

If there are any $\beta$-redices left in terms after the
$\beta$-reduction, then these terms can intuitively still be
evaluated. When we have finished all such evaluations, we may
think of having reached a final value, $\ie$ a {\it $\beta$-normal
form}.
\begin{defn}
A $\lambda$-term $t$ which contains no $\beta$-redices is called a
{\it $\beta$-normal form}.
\end{defn}

We can define the notion of equality that we desire by using the
notions of $\beta$-contraction and $\alpha$-conversion as follows.

\begin{defn}
We say a term $t$ is {\it $\beta$-convertible to} a term $s$ if
and only if $s$ is obtained from $t$ by a finite (perhaps empty)
series of $\beta$-contractions, reversed $\beta$-contractions and
$\alpha$-conversions.
\end{defn}

To qualify as a satisfactory notion of equality, the
$\beta$-convertibility relation between $\lambda$-terms must
possess the properties of being symmetric, reflexive and
transitive. Thus, it must be an equivalence relation, which is
assured by the following theorem.

\begin{theorem}
The relations $\beta$-conversion and $\alpha$-conversion are
equivalence relations.
\end{theorem}
\noindent The proof of this theorem can be found
in~\cite{Hindley86ILC}.

Now we need a method to determine equality between
$\lambda$-terms based on $\beta$-conversions. We could first try
to reduce these terms to their $\beta$-normal forms, and then
check if these normal forms are identical, allowing for renaming of
bound variables. If the $\beta$-normal forms of the terms we are
trying to compare exist, this comparison approach is justified by
the {\it Church-Rosser Theorem for $\beta$-conversion} and its
corollary.
\begin{theorem}[Church-Rosser Theorem for
$\beta$-conversion]\label{th:conversion} If a term $p$ is
$\beta$-convertible to a term $q$, then there exists a term $t$
such that $p$ and $q$ both $\beta$-reduce to $t$.
\end{theorem}
This theorem follows from the {\it Church-Rosser Theorem for
$\beta$-reduction}~\cite{Hindley86ILC} and its proof can be found
in~\cite{Hindley86ILC}.

From the {\it Church-Rosser Theorem for $\beta$-conversion}, we
can obtain the following corollary.
\begin{corollary}\label{co:conversion}
If normal forms exist for the terms $t$ and $s$, then $t$ and $s$
are $\beta$-convertible if and only if their normal forms are
identical up to $\alpha$-conversions.
\end{corollary}
\begin{proof}
By the definition of $\beta$-conversion, a term and its normal
form are $\beta$-convertible to each other. The necessity of this
claim is obvious because two identical (up to $\alpha$-conversion)
terms are $\beta$-convertible to each other and $\beta$-conversion
is transitive. The sufficiency of this claim is proved as the
following: suppose $t'$ and $s'$ are the normal forms of $t$ and
$s$ respectively. By the transitivity of $\beta$-conversion, $t'$
is $\beta$-convertible to $s'$. Theorem~\ref{th:conversion} gives
a term $r$ that $t'$ and $s'$ both reduce to. Since $t'$ and $s'$
contain no redices, they must both be $\alpha$-convertible to
$r$.
\end{proof}

Consider the comparison of two $\lambda$-terms $((\lambdax{x}\app
(x\app t_1))\app a)$ and $((\lambdax{x}\app(x\app t_2))\app b)$,
where $a$ and $b$ are distinct constants. The reductions of the
top-level redices generate two terms in the form of $(a\app
t_1[x:=a])$ and $(b\app t_2[x:=b])$. Since the difference between
constants $a$ and $b$ already implies that the normal forms of the
two terms we are trying to compare can not be identical modulo
$\alpha$-conversion, we are not interested in the reduction
results of $t_1[x:=a]$ and $t_2[x:=b]$ any more and, therefore,
the reduction on $t_1[x:=a]$ and $t_2[x:=b]$ can be ignored. To
describe such an improvement of our comparison approach, the idea
of a head normal form is useful.

\begin{defn}\label{df:GHNF}
We say a $\lambda$-term is in head normal form if it has the
structure $(\lambdax{x_1}\, \ldots\, (\lambdax{x_n}\, (\ldots
(h\app t_1) \app \ldots \app t_m))\ldots)$ where $h$ is a
constant, any one of $x_1,\ldots, x_n$ or an instantiatable
variable. By a harmless abuse of the notation, we permit $n$ and
$m$ to be $0$ in this presentation. Given such a form, $t_1,
\ldots, t_m$ are called its arguments, $h$ is called its head,
$x_1,\ldots , x_n$ are called its binders and $n$ is its binder
length.
\end{defn}
We can observe that a term has a normal form only if it has a head
normal form.

There are certain kinds of reduction sequences that are guaranteed
to produce a head normal form of a given term whenever one exists.
The following definition identifies a sequence of this kind.

\begin{defn}~\label{df:HRS-g}
The {\it head redex} of a $\lambda$-term $t$ that is not a head
normal form is identified as follows. If $t$ is a redex, then it
is its own head redex. Otherwise $t$ must be of the form $(t_1\app
t_2)$ or $(\lambdax{x}\app t_1)$. In either case, the head redex
of $t$ is identical to that of $t_1$. \ignore{The {\it weak head
redex} of a $\lambda$-term that is not a weak head normal form is
defined similarly, except that the term in question cannot be an
abstraction.}

The {\it head reduction sequence} of a term $r_0$ is
the sequence $s = r_0,r_1,r_2,\dots, r_n,\dots$, where, for $i
\geq 0$, there is a term succeeding $r_i$ if $r_i$ is not a head
normal form and, in this case, $r_{i+1}$ is obtained from $r_i$ by
rewriting the head redex using the $\beta$-contraction rule. Such
a sequence is obviously unique and terminates just in case there
is an $m \geq 0$ such that $r_m$ is a head normal form.
\end{defn}

The following theorem justifies that if a term has a head normal
form, then its head reduction sequence will terminate with such a
form.

\begin{theorem}~\label{th:HRS}
A $\lambda$-term $t$ has a head normal form if and only if the
head reduction sequence of $t$ terminates.
\end{theorem}
The proof of this theorem can be found in~\cite{Bar81}.

Thus to compare two $\lambda$-terms, we can first try to reduce
them into their head normal forms following their head reduction
sequences, and then check the identity of the binder lengths and
the number of arguments and the identity (up to
$\alpha$-conversion) of the heads of these head normal forms.
After that, we can proceed to compare their arguments,
if this is still relevant. \ignore{ It is
obvious that two head normal forms are identical (up to
$\alpha$-conversion) if and only if their binder lengths and heads
are identical (up to $\alpha$-conversion), they each have the same
number of arguments and their arguments, taken pairwise, are
equal. Further a term has a normal form only if it has a head
normal form. Thus the actual comparison method of $\lambda$-terms
usually proceeds by reducing them to their head normal forms,
matching their binders and heads and then proceeding to compare
their arguments if this is still relevant. We will follow this
method to use head normal forms to interleave comparison and
reduction processes in our later discussion on the reduction
strategies. }

As we discussed previously, this comparison method is meaningful
only if the (head) normal forms of the terms we are trying to
compare exist. There is no promise of such an
existence in the untyped $\lambda$-calculus. However, in the
context of representing syntactic objects, we are interested
eventually in typed $\lambda$-calculi in which the set of terms
is restricted to contain only the typeable ones of the untyped
$\lambda$-calculus. In most useful cases of typed
$\lambda$-calculi, head normal forms are known to exist for every
term.

\section{The De Bruijn Notation}\label{sec:debruijn}
In the comparison approach we illustrated previously, after the
$\lambda$-terms are reduced to their head normal forms, we need to
check the identity of their heads based on $\alpha$-conversions,
$\ie$ renaming the relevant bound variables. From the perspective
of real implementations, the ease of $\alpha$-conversions in the
identity checking is of special significance. The
notation proposed by de Bruijn \cite{debruijn72} provides an
elegant way of handling this problem by ignoring the bound
variable names and thus the need of renaming bound
variables during identity checking. In this section, we give
an overview of the de Bruijn notation.

\subsection{Terms in the De Bruijn Notation}\label{subsec:db-terms}
The de Bruijn terms are defined as follows.
\begin{defn}
The set of de Bruijn terms are given by the following syntactic
rules.
\begin{tabbing}
\quad\=$\langle DT\rangle$\ \=::=\ \=\kill \>$\langle
DT\rangle$\>::=\>$\langle C \rangle \ \vert\ \langle
V \rangle \ \vert  \ \#\langle I \rangle\ \vert
(\langle DT\rangle\app \langle DT\rangle)\ \vert\
(\lambdadb \langle DT\rangle)$
\end{tabbing}
In these rules, $\langle C \rangle$ represents constants, $\langle
V \rangle$ represents instantiatable variables and $\langle I
\rangle$ represents the category of positive numbers.
\end{defn}

In the de Bruijn notation, a bound variable occurrence is denoted
by an index which counts the number of abstractions between the
occurrence of this variable and the abstraction binding it.
Since we are in fact interested in only top-level closed terms,
all the abstractable variables can be transformed to their indices
in this way. This correspondence is exposed by the transformation
function $\zeta$ defined as the following, where we
correspondingly assume that the term to be transformed is closed at the
top-level.
\begin{defn}\label{def:name-deb}
The mapping $\zeta$ from the closed name-carrying terms to the de
Bruijn terms is given as $\zeta(t)=\xi(t,nil)$ where $\xi$ is a
mapping from the class of name-carrying terms and the class of
bound variable name lists to the class of de Bruijn terms and is
defined as follows. (For simplicity, we assume that the bound
variable names of the term to be transformed are distinct from each
other without loss of generality.)
\begin{enumerate}
\item If $t$ is a constant, $\xi (t,l) = t$;
\item If $t$ is an instantiatable variable, $\xi (t,l) = t$;
\item If $t$ is an abstractable variable, $\xi (t,l) = \# i$, where $t= ith(l)$;
\item If $t$ is an application $(t_1\app t_2)$,
$\xi((t_1\app t_2),l) = (\xi(t_1,l)\app\xi(t_2,l))$;
\item If $t$ is an abstraction $(\lambdax{x}\app t_1)$,
$\xi((\lambdax{x}\app t_1),l) = \lambdadb (\xi (t_1,x::l))$.
\end{enumerate}
\end{defn}

For example, consider two $\alpha$-convertible terms
\begin{tabbing}
\qquad\=\kill \> $(\lambdax{x}\app
(\lambdax{y}\app x \app y) \app x)$ and $(\lambdax{z}\app
(\lambdax{w}\app z \app w) \app z)$
\end{tabbing}
in the name-carrying scheme.
Their de Bruijn representations obtained from the application
of the transformation function $\zeta$ both have the same form as
\begin{tabbing}
\qquad\=\kill \>
$(\lambdadb(\lambdadb \# 2 \app \# 1)\app \# 1)$.
\end{tabbing}
In general, it can be seen that the need for bound variable
renaming is eliminated in determining the identity of the de
Bruijn terms.

\subsection{Substitutions in the De Bruijn Notation}\label{subsec:db-sub}
Since an index is used to count the number of abstractions between
the occurrence of a bound variable and the abstraction binding it,
extra attention should be paid when a term is substituted into the
bodies of some abstractions. Consider the term
$(\lambdadb((\lambdadb\lambdadb((\# 1\app \# 2)\app \#
3))(\lambdadb\# 2)))$. The $\beta$-contraction of the redex inside
this term requires substituting $(\lambdadb\# 2)$ for the first
free variable of $(\lambdadb((\# 1\app\# 2)\# 3))$ which is
denoted by the index $\#2$. The first thing we need to note is
that the variable occurrence represented by index $\#2$ in
$(\lambdadb\# 2)$ is in fact bound by the top level abstraction of
the entire term. Thus after the substitution, this index should be
increased by one to preserve the correct binding relation, because
there appears an extra abstraction between this variable
occurrence and the abstraction binding it. Second, the variable
occurrence represented by the index $\# 3$ in the subterm
$(\lambdadb((\# 1\app \# 2)\app \# 3))$ is also bound by the
top-level abstraction of the entire term. After the contraction,
its index should be decreased by one to reflect that an
abstraction between this variable occurrence and the abstraction
binding it disappears. In fact, all the free variable occurrences
of this subterm should be affected in this way. For this reason,
we are more interested in a generalized notation of substitutions,
that of substituting terms for all the free variables
simultaneously, the precise definition of which is given as the
following.

\begin{defn}
Let $t$ be a de Bruijn term and let $s_1$, $s_2$, $s_3$,... be an
infinite sequence of de Bruijn terms. The result of simultaneously
substituting $s_i$ for the $i$th free variable of $t$ is denoted
by $S(t;s_1,s_2,s_3,...)$ and is defined recursively as the
following:
\begin{enumerate}
\item if $t$ is a constant, $S(t;s_1,s_2,s_3,...)=t$;
\item if $t$ is an instantiatable variable, $S(t;s_1,s_2,s_3,...)=t$;
\item if $t$ is a variable reference $\# i$, $S(t;s_1,s_2,s_3,...)=s_i$;
\item if $t$ is an application $(t_1\app t_2)$, \\
$S((t_1\app t_2);s_1,s_2,s_3,...)=(S(t_1;s_1,s_2,s_3,...)S(t_2;s_1,s_2,
s_3,...))$;
\item if $t$ is an abstraction $(\lambdadb t_1)$, \\
$S((\lambdadb t_1);s_1,s_2,s_3,...)=(\lambdadb S(t_1;\# 1,s_1',s_2',s_3',...))$, \\
       where, for $i\geq 1$, $s_i'=S(s_i;\#2,\#3,\#4,...)$.
\end{enumerate}
\end{defn}

The last substitution rule is used to deal with some of the issues
we illustrated by the previous example. First, we note that within
an abstraction $(\lambdadb t)$, the first free variable has an
index $\# 2$, the second has an index $\# 3$ and so on. Since
$s_i$ is intended to be substituted for the $i$th {\it free}
variable of $t$, the variable occurrences with indices less than
$\# 2$, which are bound in $t$, should not be changed by
this substitution. 
Thus, in rule (5), we add $\#1$ in the front of the infinite
substitution sequence to achieve such a protection. Furthermore,
since an extra abstraction appears in front of $s_i$ after
the substitution is pushed into the abstraction, the indices of
the free variables of $s_i$ should be increased by one to
preserve the correct binding relation. Substitution $S(s_i; \#2,
\#3, \#4,...)$ is used for this renumbering operation. Note that
although the implicit use of $\alpha$-conversions is eliminated
from the substitution operation, extra effort should be made to
renumber the indices of the free variables to preserve the correct
binding relation.

\subsection{Rule of $\beta$-Contraction in the De Bruijn Notation}\label{subsec:db-beta}
With the presentation of substitutions in the de Bruijn notation,
we can now formally describe the $\beta$-contraction schema within
this context.

\begin{defn}
The $\beta$-contraction rule in the de Bruijn notation is the
following.
\begin{tabbing}
\qquad\=\kill \> $((\lambdadb t_1)\app t_2) \rightarrow
S(t_1;t_2,\#1,\#2,...)$,
\end{tabbing}
where $t_1$ and $t_2$ are de Bruijn terms.

Let $t$ be a de Bruijn
term that has a subterm in the form of $((\lambdadb t_1)\app t_2)$.
If the de
Bruijn term $s$ is obtained from $t$ by replacing its subterm
$((\lambdadb t_1)\app t_2)$ with $S(t_1;t_2,\#1,\#2,...)$, then
we say $t \rhd^d_{\beta} s$.
\end{defn}

The intuitive meaning of this rule is: after the contraction, the
occurrences of the first free variable of $t_1$ should be replaced
with the term $t_2$. At the same time, the indices of all the
other free variables of $t_1$ should be decreased by one to
reflect the disappearance of the abstraction in the front.

Now we want to utilize the de Bruijn notation in our comparison
approach to eliminate the renaming operation during the identity
checking. In particular, we can first translate the name-carrying
terms into their de Bruijn representations, and then follow the
method we discussed in Section~\ref{subsec:con} to compare these
de Bruijn terms. This approach is meaningful only if the
properties of the name-carrying $\lambda$-calculus we discussed in
the Section~\ref{subsec:con} still hold in the de Bruijn notation.
We first adapt Definition~\ref{df:GHNF} and~\ref{df:HRS-g} to the
de Bruijn notation in the obvious way, and then use
Theorem~\ref{th:name-deb} and~\ref{th:deb-name} to exhibit a close
correspondence between the contractions in the name-carrying
$\lambda$-calculus and those in the de Bruijn notation, and hence
assure that the properties we are interested in still follow in
the context of the de Bruijn notation.

The following two lemmas are used in the proofs of
Theorems~\ref{th:name-deb} and~\ref{th:deb-name}.

Consider the situation of substituting a name-carrying term $t$
into a context embedded under $m$ abstractions. Lemma~\ref{lemma1}
assures that the result of the transformation of $t$ is the same no
matter whether we transform $t$ to its de Bruijn representation first, and
then increase the indices of its free variables by $m$, or we
preform the substitution in the name-carrying scheme first, and
then transform the entire term.

\begin{lemma}\label{lemma1}
Let $t$ be a name-carrying term with all the names of its free
variables contained in the list $(u_1::...::u_n::l)$.
\begin{tabbing}
\qquad\= = \=\kill \> $S(\xi (t,u_1::...::u_n::l);\# 1, \#
2,...,\#n, \#(m+1+n), \# (m+2+n),...)$\\
\>\>$= \xi (t,u_1::...::u_n::v_1::v_2::...::v_m::l)$,
\end{tabbing}
where each $v_i$, for $1\leq i\leq m$, does not occur amongst
$u_1$,...,$u_n$, or in $l$.
\end{lemma}
This lemma can be proved by straightforward induction on the
structure of $t$.

Consider the situation of substituting a name-carrying term $t_2$
for the free variable $x$ in a name-carrying term $t_1$.
Lemma~\ref{lemma2} assures that the result of transformation is the
same no matter whether we perform the substitution $t_1[x:=t_2]$ first,
and then translate the entire term to its de Bruijn
representation, or we transform $t_1$ and $t_2$ first, and then
perform the substitution in the context of de Bruijn notation.

\begin{lemma}\label{lemma2}
Let $t_1$ be a name-carrying term with all the names of its free
variables contained in the list $(v_1::v_2::...::v_m::x::l)$.
\begin{tabbing}
\qquad\= = \= \kill \>$S(\xi (t_1, v_1::v_2::...::v_m::x::l);\#
1,...,\# m, s', \#
(m+1), \# (m+2),...) $\\
\>\>$= \xi(t_1[x:=t_2], v_1::v_2::...::v_m::l)$,
\end{tabbing}
where $s' = S(\xi(t_2,l); \# (m+1), \# (m+2), ... )$.
\end{lemma}
With the aid of Lemma~\ref{lemma1}, this lemma can be easily proved
by induction on the structure of $t_1$.

\begin{theorem}\label{th:name-deb}
Let $t_1$ be a name-carrying $\lambda$-term such that all the
names of its free variables are contained in the list $l$. If $t_1
\rhd_{\beta} t_2$, then $\xi(t_1,l) \rhd^d_{\beta} \xi(t_2,l)$.
Further, if the contracted redex is a head redex in the
name-carrying scheme, then the translation to the de Bruijn term is
realized by contracting a head redex in the de Bruijn notation.
\end{theorem}
\begin{proof}
The first part of this theorem is proved by induction on the
structure of $t_1$, and the second part can be easily observed
during this process.

Now we consider the cases of the structure
of $t_1$.
Since $t_1$ contains a redex, it can only be an abstraction or an
application.

Suppose $t_1$ is an application $(s_1\app s_2)$. There are two
subcases: first, $t_1$ itself is the redex contracted; second, the
redex contracted is a subterm of one of $s_1$ or $s_2$. In the
first case, as a redex being contracted, $t_1$ is in the form
of $((\lambdax{x}\app s'_1)\app s_2)$, and correspondingly, $t_2$
is in the form of $s'_1[x:=s_2]$. According to the definition of
function $\xi$,
\begin{tabbing}
\qquad\=\kill \>
$\xi(t_1,l)=(\lambdadb\xi(s'_1,x::l))\app\xi(s_2,l)$ and
$\xi(t_2,l)=\xi(s'_1[x:=s_2],l)$.
\end{tabbing}
Following the $\beta$-contraction rule in the de Bruijn notation,
\begin{tabbing}
\qquad\=\kill \>
$\xi(t_1,l)\rhd^d_{\beta}S(\xi(s'_1,x::l);\xi(s_2,l),\#1,\#2,...)$.
\end{tabbing}
As an instance of Lemma~\ref{lemma2},
\begin{tabbing}
\qquad\=\kill \>
$S(\xi(s'_1,x::l);\xi(s_2,l),\#1,\#2,...)=\xi(s'_1[x:=s_2],l)$.
\end{tabbing}
In the second case, without loss of generality, we assume that
the redex contracted is a subterm of $s_1$. Thus $t_2$ must have
the form of $(s'_1\app s_2)$, where $s_1\rhd_{\beta} s'_1$. By Lemma
\ref{free variables}, we know that $l$ contains all
the free variable names of $s'_1$.
By the
induction hypothesis, $\xi(s_1,l)\rhd^d_{\beta}\xi(s'_1,l)$.
Clearly,
\begin{tabbing}
\qquad\=\kill \> $(\xi(s_1,l)\app\xi(s_2,l))
\rhd^d_{\beta}(\xi(s'_1,l)\app\xi(s_2,l))$.
\end{tabbing}
According to the definition of function $\xi$,
\begin{tabbing}
\qquad\=\kill \> $\xi(t_1,l)=(\xi(s_1,l)\app \xi(s_2,l))$ and
$\xi(t_2,l)=(\xi(s'_1,l)\app \xi(s_2,l))$.
\end{tabbing}
Thus we have proven that $\xi(t_1,l)\rhd^d_{\beta}\xi(t_2,l)$ in the
case that $t_1$ is an application.

Suppose $t_1$ is an abstraction $(\lambdax{x}\app t'_1)$. Since
$t_1\rhd_{\beta} t_2$, $t_2$ must be in the form of $(\lambdax{x}\app
t'_2)$ where $t'_1\rhd_{\beta} t'_2$. By Lemma \ref{free variables},
$FV(t'_2)\in FV(t'_1)$ and therefore, all the free variable names
of $t_1$ and $t_2$ are contained in the list $(x::l)$. Hence, by the
induction hypothesis,
\begin{tabbing}
\qquad\=\kill \> $\xi(t'_1,x::l) \rhd^d_{\beta}\xi(t'_2,x::l)$.
\end{tabbing}
Hence
\begin{tabbing}
\qquad\=\kill \> $(\lambdadb
\xi(t'_1,x::l))\rhd^d_{\beta}(\lambdadb\xi(t'_2,x::l))$.
\end{tabbing}
According to the definition of function $\xi$,
\begin{tabbing}
\qquad\=\kill \> $\xi(t_1,l)=(\lambdadb \xi(t'_1,x::l))$ and
$\xi(t_2,l)=(\lambdadb\xi(t'_2,x::l))$.
\end{tabbing}
Thus we have proven that $\xi(t_1,l)\rhd^d_{\beta}\xi(t_2,l)$ in this
case.
\end{proof}

\begin{theorem}\label{th:deb-name}
Let $t_1$ be a name-carrying $\lambda$-term such that all the
names of its free variables are contained in the list $l$. If
$\xi(t_1,l) \rhd^d_{\beta} \xi(t_2,l)$, then $t_1 \rhd_{\beta}
t_2$. Further, if the contracted redex
is a head redex in the de Bruijn notation, then
the corresponding redex before the transformation is a head redex
in the name-carrying scheme.
\end{theorem}
\begin{proof}
The first part of this theorem is proved by induction on the
structure of $t_1$, and the second part can be easily observed
during this process.

Now we consider the cases of the structures of $t$.
Since $\xi(t_1,l)$ contains a redex, $\xi(t_1,l)$ can only be an
application or an abstraction and therefore so can $t_1$.

Suppose $t_1$ is an application $(s_1\app s_2)$. Then
\begin{tabbing}
\qquad\=\kill \> $\xi(t_1,l)= (\xi(s_1,l)\app\xi(s_2,l))$.
\end{tabbing}
If the redex contracted is $\xi(t_1,l)$ itself, then
$\xi(s_1,l)$ must have the form $\lambdadb \xi(s'_1,x::l)$, and
$\xi(t_2,l)$ must be in the form
\begin{tabbing}
\qquad\=\kill \> $S(\xi(s'_1,x::l);\xi(s_2,l),\#1,\#2,...)$.
\end{tabbing}
According to the definition of function $\xi$,
\begin{tabbing}
\qquad\=\kill \> $\lambdadb \xi(s'_1,x::l) = \xi(\lambdax{x}\app
s'_1,l)$.
\end{tabbing}
Thus $t_1$ has the form $((\lambdax{x}\app s'_1)\app s_2)$.
Following the $\beta$-contraction rule in the name-carrying
scheme, this redex should be rewritten to $s'_1[x:=s_2]$. As an
instance of Lemma~\ref{lemma2},
\begin{tabbing}
\qquad\=\kill \>
$S(\xi(s'_1,x::l);\xi(s_2,l),\#1,\#2,...)=\xi(s'_1[x:=s_2])$.
\end{tabbing}
Now we consider the case that the redex contracted is not
$\xi(t_1,l)$ itself but is a subterm of one of $\xi(s_1,l)$ or
$\xi(s_2,l)$. Without loss of generality, we assume the redex
contracted is inside $\xi(s_1,l)$, and
$\xi(s_1,l)\rhd^d_{\beta}\xi(s'_1,l)$. Thus $\xi(t_2,l)$ must be in the
form
\begin{tabbing}
\qquad\=\kill \> $(\xi(s'_1,l)\app\xi(s_2,l))$,
\end{tabbing}
and therefore $t_2=(s'_1\app s_2)$. By the induction hypothesis,
$s_1\rhd_{\beta} s'_1$ in the name carrying scheme. Clearly,
$(s_1\app s_2)\rhd_{\beta} (s'_1\app s_2)$. Thus we have proven that
$t_1\rhd_{\beta}t_2$ in the case that $t_1$ is an application.

Suppose $t_1$ is an abstraction $(\lambdax{x}\app t'_1)$. Then
\begin{tabbing}
\qquad\=\kill \> $\xi(t_1,l)=\lambdadb \xi(t'_1,x::l)$.
\end{tabbing}
Since
 $\xi(t_1,l)\rhd^d_{\beta}\xi(t_2,l)$,
$\xi(t_2,l)$ must have the form  $\lambdadb
\xi(t'_2,x::l)$, where
\begin{tabbing}
\qquad\=\kill \> $\xi(t'_1,x::l)\rhd^d_{\beta}\xi(t'_2,x::l)$.
\end{tabbing}
Thus, $t_2$ has the form of $(\lambdax{x} \app t'_2)$. By the
induction hypothesis, $t'_1\rhd_{\beta} t'_2$ in the name-carrying
scheme. Clearly, $(\lambdax{x}\app
t'_1)\rhd_{\beta}(\lambdax{x}\app t'_2)$. Thus we have proven that
$t_1\rhd_{\beta}t_2$ in this case.
\end{proof}

Suppose two terms $t$ and $s$ are $\beta$-convertible to each
other in the name-carrying scheme. Then $\zeta(t)$ and $\zeta(s)$
are also $\beta$-convertible in the context of the de Bruijn
notation. By Corollary~\ref{co:conversion}, we know that $t$ and
$s$ have a common normal form $p$ (up to $\alpha$-conversion), if
their normal forms exist. According to Theorem~\ref{th:name-deb}
and Theorem~\ref{th:deb-name}, the contractions performed on $t$
and $s$ will be mapped exactly to those performed on $\zeta(t)$
and $\zeta(s)$. Thus $\zeta(t)$ and $\zeta(s)$ have a common
(head) normal form $\zeta(p)$, if the normal forms of $\zeta(t)$
and $\zeta(s)$ exist. Further, Theorem~\ref{th:name-deb} and
Theorem~\ref{th:deb-name} also assure that the contraction steps
performed on a name-carrying term $t$ when following its head reduction
sequence will be mapped exactly to those performed on $\zeta(t)$
when following its head reduction sequence in the context of the de
Bruijn notation. Therefore Theorem~\ref{th:HRS} holds in this
context, too. Finally, we are only interested in the terms
for which normal forms exist. Thus
we can refine our comparison approach
to the following: we first transform the name-carrying terms under
comparison into their de Bruijn representations, and then try to
reduce these de Bruijn terms into their head normal forms. After
that, we simply match the binder lengths and the heads of those
head normal forms, and proceed to compare their arguments if this
is still relevant. It is clear that the main issue of the
comparison approach is the head reduction process of the de Bruijn
terms under comparison.

\ignore{ According to the close correspondence between the
contraction steps in the de Bruijn notation and the name-carrying
representation, it is not hard to see that the {\it Church-Rosser
Theorem of $\beta$-conversion} and its corollary also hold in the
context of the de Bruijn notation. Further, adapting
Definition~\ref{df:GHNF} and~\ref{df:HRS-g} to the de Bruijn
notation, it is obvious that a de Bruijn term has a normal only if
it has a head normal form and Theorem~\ref{th:HRS} also holds in
this context. Finally, since we are only interested in the terms
having their normal forms, the approach we discussed in the
previous section still works in this context. Thus to compare two
name-carrying $\lambda$-terms, we can first transform them into
their de Bruijn representations, then try to reduce these de
Bruijn terms into their head normal forms, after that we simply
match the binder lengths and the heads of those head normal forms
and, finally, proceed to compare their arguments. It can be easily
seen that the main issue of this comparison approach is the
reduction operation from the de Bruijn terms to their head normal
forms. }

\section{Explicit Substitution Calculi}\label{sec:exp-sub}
Hitherto the substitution in the course of $\beta$-reduction has
still been viewed as a rather atomic operation in the sense that
once generated, it will be performed on the corresponding term
structures immediately. However, the performance of substitutions
consists of the traversals of the term structures which are not
yet, but will be, reduced. Thus, if we can temporarily suspend the
substitutions once they are generated and delay their performance
so that they can be carried out alone with the reduction steps,
the substitution process can gain the ability to interact more
with the reduction process. For instance, we can hopefully combine
the substitutions generated at different reduction stages and
which are to be performed on the same term structures, and carry
them out in one term traversal. Thus we can also avoid some
redundant effort incurred by the performance of those
substitutions which turns out to be unnecessary due to later
reduction results.

Explicit substitution calculi extend the de Bruijn notation to
record suspended substitutions directly into term structures,
thereby offering our desired flexibility in ordering computations.
We study the benefits of this flexibility in this thesis based on
a particular calculus known as the {\it suspension notation}
\cite{NW98tcs}. We outline this notation in this section to
facilitate this discussion. Although our empirical study must
utilize a particular system, the suspension notation is general
enough for our observations to eventually be calculus independent.
We make this point below by contrasting this system with the other
explicit substitution calculi in existence.

\subsection{The Suspension Notation}\label{subsec:suspension}
To explicitly record substitutions, the explicit substitution
notations involve at least two syntactic categories: those
correspond to terms, and to the environment in which the suspended
substitutions are recorded. As we noticed in
Section~\ref{subsec:db-sub}, when substitutions are pushed into an
abstraction, the free variable indices of terms to be substituted
in should be increased by one to reflect the change in their
embedding levels. Thus in a notation such as the
$\lambdasi$-calculus~\cite{ACCL91} that uses exactly two
categories of expressions, such an adjustment should be performed
on the entire environment each time that the environment is
propagated into an abstraction. The suspension notation instead
uses a global mechanism for recording the adjustment to be made on
the free variable indices of the terms to be substituted in, so
that the adjustment can be made only once at the time that the
substitutions are actually performed, rather than in an iterated
manner. To support this possibility, the suspension notation
includes a third category of expressions called environment terms
that encode terms to be substituted in, together with their
embedding context.

\begin{defn}\label{def:suspension}
The syntactical definition of {\it suspension terms} is given by
the following rules.
\begin{tabbing}
\quad\=$\langle STerm\rangle$\ \=::=\ \=\kill \>$\langle
STerm\rangle$\>::=\>$\langle C \rangle \ \vert\ \langle V \rangle
\ \vert  \ \#\langle I \rangle\ \vert\
(\langle STerm\rangle\app \langle STerm\rangle)\ \vert\ $ \\
\>\>\> $(\lambdadb
\langle STerm\rangle)\ \vert \ \env{\langle STerm\rangle,\langle N \rangle, \langle N \rangle, \langle Env \rangle}$\\
\>$\langle Env \rangle$\> ::= \> $nil\ \vert\ \langle
ETerm \rangle :: \langle Env \rangle$\\
\>$\langle ETerm \rangle$\> ::= \> $\dum \langle N
\rangle\ \vert\ (\langle STerm \rangle, \langle N \rangle)$\\
\end{tabbing}
In these rules, $\langle C \rangle$ represents constants, $\langle
V \rangle $ represents instantiatable variables, $\langle I
\rangle$ is the category of positive numbers and $\langle N
\rangle$ is the category of nonnegative numbers.
\end{defn}

Besides the de Bruijn terms, there is a new type of terms called
$\emph{suspensions}$, of the form $\env{t,ol,nl,e}$,
corresponding to the temporarily suspended
substitutions with the term they should be performed on. The
intuitive meaning of a suspension in this form
is that the first $ol$ variables of term $t$ should be substituted
for in a way determined by $e$ and the other variables of $t$
should be renumbered to reflect the fact that the embedding
level of $t$ was originally $ol$ but now is $nl$. Note that a
suspension has the ability to record multiple substitutions
generated from the contractions of different redices. This ability
is necessary for the realization of substitution combinations.

An environment, corresponding to the category $\langle Env \rangle$, is a
finite list in which the term to be substituted for the $i$th free
variable, together with its embedding context, is maintained in the
$i$th position. Hence in a well-formed suspension term, the length
of this environment list must be the same as $ol$.

Represented by the category $\langle ETerm \rangle$, two kinds of
environment terms can appear within the environment. They correspond 
to variables bound by two different types of abstractions in the 
original term: $(t,l)$ denotes a term replacement to be made on the 
variables bound by abstractions which disappear after reductions, while
$\dum l$ represents the adjustment to be made on the variables bound 
by abstractions which persist.  
The natural number $l$
inside these two kinds of terms encodes the new embedding level at
the relevant abstraction, $\ie$ for the variables bound by                   
abstractions that persist after the reduction, we intend this to
be the new embedding level just within the scope of the
abstraction. 
Consequently, there are also
constraints on $nl$ and $l$ in a well-formed suspension term: the
$l$ in $(t,l)$ should be less than or equal to $nl$, and the $l$ in
$\dum l$ should be less than $nl$.

\begin{figure}[!ht]
\begin{tabbing}
\quad\=(r11)\ \=\qquad\qquad\qquad\=\kill \> ($\beta_s$)\>
$((\lambdadb t_1)\app t_2) \ra \env{ t_1, 1,
0, (t_2,0) :: nil}$\\
\> ($\beta'_s$)\> $((\lambdadb
\env{t_1,ol+1,nl+1,\dum{nl}::e})\app t_2) \ra  \env{ t_1,ol+1,nl, (t_2,nl) :: e }$\\
\>(r1)\> $\env{c,ol,nl,e} \ra c$\\
\>\> provided $c$ is a constant\\
\>(r2)\> $\env{x,ol,nl,e} \ra x$\\
\>\>provided $x$ is an instantiatable
 variable\\
\>(r3)\>$\env{\#i,ol,nl,e} \ra \#j$\\
\>\>provided $i > ol$ and $j = i - ol + nl$.\\
\>(r4)\>$\env{\#i,ol,nl, e} \ra \#j$\\
\>\>provided $i \leq ol$ and $e[i] = \dum{l}$ and $j = nl - l$.\\
\>(r5)\>$\env{\#i,ol,nl,e} \ra \env{t,0,j,nil}$\\
\>\> provided $i \leq ol$ and $e[i] = (t,l)$ and $j = nl - l$.\\
 \> (r6)\> $\env{(t_1\app t_2),ol,nl,e} \ra
(\env{t_1,ol,nl,e}\app \env{t_2,ol,nl,e})$.\\
\>(r7)\>$\env{(\lambdadb t), ol, nl, e} \ra (\lambdadb
\env{t, ol+1, nl+1, \dum{nl} :: e})$.\\
\>(r8)\>$\env{\env{t,ol,nl,e},0,nl',nil} \ra \env{t,ol,nl+nl',e}.$ \\
\>(r9)\>$\env{t,0,0,nil} \ra t$
\end{tabbing}
\caption{Rewrite rules for the suspension notation}
\label{fig:rewriterules}
\end{figure}

Along with term representations, there is a collection of rewrite
rules to simulate $\beta$-reduction. These rules are presented in
Figure~\ref{fig:rewriterules}. We use $e[i]$ to refer to the $i$th
item in the environment list.

Among these rules, $\beta _s$ and $\beta' _s$ generate the
suspended substitutions corresponding to $\beta$-contraction;
rules (r1)-(r9), referred to as \emph{reading rules}, are used to
actually carry out those substitutions.

Now we use a concrete example to illustrate the roles played by
the rewrite rules to simulate $\beta$-reduction and the way in
which the substitutions are combined. Consider the de Bruijn term
\begin{tabbing}
\qquad\=\kill \> $((\lambdadb ((\lambdadb (\lambdadb ((\#1\app
\#2)\app \#3))) \app t_2))\app t_3)$,
\end{tabbing}
where $t_2$ and $t_3$ are arbitrary de Bruijn terms. Using rule
($\beta_s$) to contract the outermost redex, the term is rewritten
to
\begin{tabbing}
\qquad\=\kill \> $\env{((\lambdadb (\lambdadb ((\#1\app \#2)\app
\#3))) \app t_2),1,0,(t_3,0)::nil}$.
\end{tabbing}
Using rule (r6) to propagate the substitution into the top-level
application inside the suspension, the term is rewritten to
\begin{tabbing}
\qquad\=\kill \> $\env{(\lambdadb (\lambdadb ((\#1\app \#2)\app
\#3))),1,0, (t_3,0)::nil}\app \env{t_2,1,0,(t_3,0)::nil}$.
\end{tabbing}
Using rule (r7) to propagate the substitution into the top-level
abstraction inside the former suspension, the whole term is rewritten to
\begin{tabbing}
\qquad\=\kill \> $(\lambdadb \env{(\lambdadb ((\#1\app \#2)\app
\#3)),2,1, \dum 0::(t_3,0)::nil})\app \env{t_2,1,0,(t_3,0)::nil}$.
\end{tabbing}
Now using rule ($\beta'_s$) to contract the redex and combine the
substitution generated by this contraction with the one already
existing in the environment, the term is rewritten to
\begin{tabbing}
\qquad\=\kill \> $\env{(\lambdadb ((\#1\app \#2)\app \#3)),2,0,
(\env{t_2,1,0,(t_3,0)::nil},0)::(t_3,0)::nil}$.
\end{tabbing}
Using rule (r7) and (r6) several times to propagate the
substitution into applications and abstractions, the term is
transformed to
\begin{tabbing}
\qquad\=\kill \>$(\lambdadb ($\=$($\=$\env{\#1,3,1,\dum 0::
(\env{t_2,1,0,(t_3,0)::nil},0)::(t_3,0)::nil}\app $ \\ \>
\>\>$\env{\#2,3,1,\dum 0::(\env{t_2,1,0,(t_3,0)::nil},0)::(t_3,0)::nil}) \app$\\
\>\> $\env{\#3,3,1,\dum 0::(\env{t_2,1,0,(t_3,0)::nil},0)::(t_3,0)::nil}))$.
\end{tabbing}
At this time, reaching the abstractable variables,
substitutions can actually be performed. Using
rule (r4) to rewrite the first suspension, the term is rewritten
to:
\begin{tabbing}
\qquad\=\kill \>$(\lambdadb ($\=$(\#1 \app
\env{\#2,3,1,\dum 0::(\env{t_2,1,0,(t_3,0)::nil},0)::(t_3,0)::nil}) \app $\\
\>\>$\env{\#3,3,1,\dum 0::(\env{t_2,1,0,(t_3,0)::nil},0)::(t_3,0)::nil}))$.
\end{tabbing}
Using rule (r5) to rewrite the current first suspension,
the term is transformed to
\begin{tabbing}
\qquad\=\kill \>$(\lambdadb ($\=$(\#1 \app
\env{\env{t_2,1,0,(t_3,0)::nil},0,
1,nil})\app $\\
\>\>$\env{\#3,3,1,\dum 0::(\env{t_2,1,0,(t_3,0)::nil},0)::(t_3,0)::nil}))$.
\end{tabbing}
Using rule (r8) to combine renumbering with the existing
substitution, the term is rewritten to
\begin{tabbing}
\qquad\=\kill \>$(\lambdadb ($\=$(\#1 \app \env{t_2,1,1,(t_3,0)::nil})\app $\\
\>\>$\env{\#3,3,1,\dum 0::(\env{t_2,1,0,(t_3,0)::nil},0)::(t_3,0)::nil}))$.
\end{tabbing}

Similarly, by the application of rule (r5), 
 the term is transformed to

\begin{tabbing}
\qquad\=\kill \>$(\lambdadb ((\#1 \app
\env{t_2,1,1,(t_3,0)::nil})\app \env{t_3,0,1,nil}))$.
\end{tabbing}
Depending on the particular structures of $t_2$ and $t_3$, the
rewrite rules can be applied to finally produce a de Bruijn term
which is $\beta$-reduced from the original term.

\begin{figure}[!h]
\begin{tabbing}
\quad\=(r11)\ \=\qquad\qquad\qquad\=\kill
\>(r10)\> $\env{\# i,ol,nl,e} \rightarrow t$, \\
\>\>  provided $i \leq ol$, $e[i]=(t,l)$ and $nl = l$. \\
\>(r11)\> $\env{\# i,ol,nl,e} \rightarrow \env{t,ol',nl'+nl-l,e'}$, \\
\>\>  provided $i \leq ol$, $e[i]=(\env{t,ol',nl',e'},l$), and $nl
\neq l$. \\
\>(r12)\> $\env{\# i,ol,nl,e} \rightarrow \env{t,0,nl-l,nil}$, \\
\>\> provided $i \leq ol$, $e[i]=(t,l)$, $t$ is not a suspension,
and $nl \neq l$.
\end{tabbing}
\caption{The enhanced version of rule (r5)}
\label{fig:enhanced-r5}
\end{figure}
If our sole purpose is to simulate $\beta$-reduction, the rule
($\beta' _s$) is redundant: whenever ($\beta'_s$) is applied, rule
($\beta_s$) is applicable too. However, as illustrated in the
previous example, ($\beta'_s$) is the rule in our rewriting system
that serves to combine the substitutions newly generated by a contraction,
with
those already recorded in the environment. This rule
requires the redex to be contracted to have the form of
\begin{tabbing}
\qquad\=\kill\> $((\lambdadb\env{t_1,ol+1,nl+1,@nl::e})\app t_2)$,
\end{tabbing}
which means that the suspension as the abstraction body is
obtained from pushing the suspension $\env{\lambdadb t_1,ol,nl,e}$
into the top-level abstraction inside it. If the reduction process
strictly follows the outermost and leftmost order, which fits the
reduction order required by head reduction sequences, all
substitutions generated during the reduction process can be
combined in this way. Similarly, rule (r8) is redundant, but serves
to combine the renumbering needed after a term has been
substituted into a new embedding context with the already existing
substitutions to be performed on the same term in the environment.
It requires the nested suspension to have the form of
$\env{\env{t,ol,nl,e},0,nl',nil}$ which means that this suspension
is generated from the application of the rule (r5). In fact, the
main uses of (r9) also arise after a use of (r5). Thus we can
further eliminate rules (r8) and (r9) in favor of the enhanced
versions of (r5) shown in Figure~\ref{fig:enhanced-r5}.

This course is followed in our reduction procedures.

\subsection{Some Formal Properties}\label{subsec:formal-prop}
\ignore{ The capacity of the suspension notation to simulate
reduction in the $\lambda$-calculus is shown in the following two
steps. First, underlying every term in the suspension notation is
intended to be a de Bruijn term that is obtained by ``calculating
out'' the suspended substitutions using the reading rules in
Figure~\ref{fig:rewriterules}. This is supported by the following
two theorems with regard to the properties of the reading rules,
the proofs of which can be found in~\cite{NW98tcs}.}

To justify that the comparison approach we discussed before still
works in the context of the suspension notation, we need to first
show that the suspension notation is capable of simulating
reductions in the de Bruijn notation.

Theorem \ref{th:readingrules-confluent} assures that for every
well-formed suspension term, there is a unique de Bruijn term underlying
it.

\begin{theorem}\label{th:readingrules-confluent}
Let $t$ be a well-formed term in the suspension notation. If terms
$t_1$ and $t_2$ are different suspension terms
obtained from $t$ by a series (maybe empty) of applications of the
reading rules, then there exists a de Bruijn term $s$, which can
be obtained from $t$ by a series (maybe empty) of applications of
the reading rules, such that $t_1$ and $t_2$ can be transformed to
$s$ by a series (maybe empty) of applications of reading rules.
\end{theorem}

Theorem \ref{th:readingrules-strongnormalizing} assures that every
rewrite sequence from a well-formed suspension term to the de Bruijn
term underlying it terminates.

\ignore{ Theorem \ref{th:readingrules-strongnormalizing} assures
that every well-formed suspension term can be transformed to the
unique de Bruijn term underlying it following whichever possible
sequence of applications of the reading rules. }

\begin{theorem}\label{th:readingrules-strongnormalizing}
Every well-formed term $t$ in the suspension notation can be
transformed to a de Bruijn term by a finite series (maybe empty)
applications of the reading rules, regardless of the specific
choice of a reading rule when there are multiple rules applicable.
\end{theorem}
The proofs of the above two theorems can be found
in~\cite{NW98tcs}.


The following theorem, which is proved in \cite{NW98tcs},
establishes the correspondence between the
reductions in the de Bruijn notation and the term transformations
in the suspension notation which are intended to simulate those
reductions.
\begin{theorem}\label{th:susp-deb}
Let $t$ be a de Bruijn term. Then $t$ $\beta$-reduces to the de Bruijn
term $s$ if and only if $t$ can be transformed to $s$ by a series
(maybe empty) of applications of rules in Figure~\ref{fig:rewriterules}
and~\ref{fig:enhanced-r5}.
\end{theorem}

\ignore{ Now in the second step, it can be shown that the de
Bruijn term $t$ $\beta$-reduces to $s$ if and only if $t$ can be
rewritten to $s$ using rewrite rules in
Figure~\ref{fig:rewriterules}. As a particular case of this, each
$\beta$-contraction in the conventional setting can be realized by
a use of the ($\beta _s$) rule followed by a sequence of
application of reading rules.}

Head normal forms are extended to the suspension notation by
permitting their arguments to be arbitrary suspension terms. For
the convenience of our later discussion, we refer to a term in the
suspension notation as being a weak head normal formal form if it
is a head normal form or it is of the form of $(\lambdadb s)$,
where $s$ is a suspension term. \ignore{In our comparison
approach, the extension of head normal forms to the suspension
notation permits the performance of those substitutions over the
arguments to be delayed until we actually need to compare the
arguments. The legitimacy of our comparison approach is based on
the following theorem that is proved in~\cite{NW98tcs}.} Following
the theorems above, Theorem \ref{th:hnfcorresp},
which is proved in~\cite{NW98tcs}, assures that
if a de Bruijn term $t$ has a (weak) head normal form $s$ in the
context of the de Bruijn notation, then $t$ has one or more (weak)
head normal forms in the suspension notation from which $s$ can be
calculated out by a finite series (maybe empty) of applications of
the reading rules.

\begin{theorem}\label{th:hnfcorresp} Let $t$ be a de Bruijn term
and suppose that the rules in Figure~\ref{fig:rewriterules} allow
$t$ to be rewritten to a (weak) head normal form in the
suspension notation that has $h$ as its head, $n$ as its binder
length and $t_1,\ldots ,t_m$ as its arguments. Let $\rnf{t_i}$ be
the de Bruijn term obtained from $t_i$ by a series (maybe empty) of
applications of reading rules. Then $t$ has the term  $$(\lambdadb
\ldots( \lambdadb (\ldots (h \app \rnf{t_1}) \app \ldots\app
\rnf{t_m})) \ldots)$$ with a binder length of $n$ as a (weak) head
normal form in the de Bruijn notation.
\end{theorem}

Comparing with the de Bruijn notation, there is one more
possibility for terms and there is also a larger set of rewriting
rules in the suspension notation. Taking these aspects into
account,
we generalize the notions of head redex and head reduction
sequence to the suspension notation and also define the notions of
weak head redex and weak head reduction sequence as the
following.
\begin{defn}\label{def:susp-hrs}
Let $t$ be a suspension term that is not in (weak) head normal
form.
\begin{enumerate}
\item Suppose that $t$ has the form $(t_1\app t_2)$.
 If $t_1$ is an abstraction, then $t$ is its sole (weak) head
redex. Otherwise the (weak) head redices of $t$ are the weak head
redices of $t_1$; notice that $t_1$ cannot be a weak head normal
form here.

\item If $t$ is of the form $(\lambdadb t_1)$,
its head redices are identical to those of $t_1$. (This case does
not arise if $t$ is not a weak head normal form.)

\item If $t$ is of the form $\env{t_1,ol,nl,e}$,
then its (weak) head redices are itself and all the (weak) head
redices of $t_1$.
\end{enumerate}
Let two subterms of a term be considered non-overlapping just in
case neither is contained in the other. Then a (weak) head
reduction sequence of a suspension term $t$ is a sequence $t =
r_0,r_1,r_2,\ldots,r_n,\ldots,$ in which, for $i \geq 0$, there is
a term succeeding $r_i$ if $r_i$ is not in (weak) head normal form
and, in this case, $r_{i+1}$ is obtained from $r_i$ by
simultaneously rewriting a finite set of non-overlapping subterms
that includes a (weak) head redex using the rule schemata in
Figure~\ref{fig:rewriterules} or \ref{fig:enhanced-r5}. Obviously,
such a sequence terminates if for some $m \geq 0$ it is the case
that $r_m$ is in (weak) head normal form.
\end{defn}

The following theorem, which is proved in~\cite{NW98tcs},
assures that if a term in the de Bruijn notation has a (weak)
head normal form then its (weak) head reduction sequences
terminate.

\begin{theorem}\label{th:HNF}
A term $t$ in the suspension notation has a (weak) head normal
form if and only if every (weak) head reduction sequence of $t$
terminates.
\end{theorem}

Thus we show that the comparison approach we illustrated before
still works in the context of the suspension notation. Further,
the extension of head normal forms to the suspension notation
permits the performance of those substitutions over the arguments
to be delayed until we actually need to compare the arguments.

\subsection{Other Explicit Substitution Calculi}\label{subsec:other-exp-sub}
According to their combination ability, explicit substitution
calculi can be classified into two categories. The calculi in the
first category do not have the ability to combine substitutions at
all. Their purposes are only to delay substitutions in the course
of simulating the $\beta$-reduction. The $\lambdaup$-calculus
\cite{BBLR96} and the $\lambdase$-calculus \cite{KR97} are two
representative calculi in this category. On the other hand, the
calculi in the second category have the ability to combine
substitutions during the reduction process. The suspension
notation we presented previously and the $\lambdasi$-calculus
\cite{ACCL91} both belong to this category.

Without the ability to combine substitutions, the terms used to
record explicit substitutions in the first kind of calculi have
the characteristic that they can each record the substitutions
generated by only one contraction. Certainly, this kind of
information can be covered by a subset of the suspension terms
with a certain pattern. For example, the two kinds of terms used
to record substitutions in the $\lambdase$-calculus can be represented
by suspension terms of the form
\begin{tabbing}
\qquad\=\kill\> $\env{t,j,j-1,\dum (j-2)::\dum (j-3)::...\dum
0::(t',0)}$,
\end{tabbing}
for renumbering after a term is substituted into a new embedding
context, and of the form
\begin{tabbing}
\qquad\=\kill\> $\env{t,k,(i-1)+k,\dum (i-2+k)::\dum
(i-3+k)::...::\dum (i-1)::nil}$,
\end{tabbing}
for the term replacement and renumbering caused by one contraction.
Correspondingly, the effects of the rewrite rules in such
calculi can also be achieved by a subset of the rewrite rules of the
suspension notation. In particular, since the rewrite rules in
those calculi are used to purely simulate the
$\beta$-contractions, the same effect can be achieved by the
rewrite rules in the suspension notation without ($\beta'_s$) and
(r8).

The other explicit substitution calculus $\lambdasi$ also
has the ability to combine substitutions to be performed on the
same term structures. 
While the $\lambdasi$-calculus has the ability to combine
arbitrarily nested suspended substitutions, in the suspension
notation we presented previously, the substitution combination
rules ($\beta'_s$) and (r8) have certain requirements on the
substitutions to be combined. Although we know that if strictly
following the leftmost and outermost reduction order, all the
substitutions can be combined by these two rules, the permission
to share the reduction results and the binding of instantiatable
variables caused by unification could sometimes violate this
reduction order and may cause the failure of the combination of
the suspended substitutions. For example, the binding of the
instantiatable variable $F$ to a suspension
$\env{t,1,0,(s,0)::nil}$ in the term $((\lambdadb F)\app t_1)$
generates the term $((\lambdadb\env{t,1,0,(s,0)::nil})\app t_1)$.
To contract this redex, the only applicable rule is ($\beta_s$).
This contraction results in a suspension term in the form of
$\env{\env{t,1,0,(s,0)::nil},1,0,(t_1,0)::nil}$. In this nested
suspension, the combination of the common substitution walks over
$t_1$ is lost. In fact, the suspension notation we represented in
this thesis is a restricted version of the calculus presented
in~\cite{NM99cade}. In particular, the full calculus allows for
the transformation from an arbitrarily nested suspension in the form
of $\env{\env{t,ol1,nl1,e1},ol2,nl2,e2}$ to a single suspension
$\env{t,ol,nl,e}$. The main task in this transformation is the
computation of the effect of the substitutions embodied in the
environment $e2$ on each of the terms in $e1$. The richer calculus
includes expression forms and rules that allow for this
computation to be carried out through genuinely atomic steps.
However, from the perspective of implementations, it is desired
that substitution combination be realized in a simple step, as
opposed to a series of operations. Secondly, in reality, following
head reduction sequences, a situation where the application of
($\beta'_s$) fails when substitution combination is needed, occurs
relatively rarely. Thus we sacrifice some of the ability to combine
substitutions
in order to simplify the combination process by using ($\beta'_s$) and (r8)
to ``over-look'' some steps of combinations and directly generate
the combination result.\ignore{The rules ($\beta'_s$) and (r8)
are designed for this purpose. 
these two rules ``over look'' some steps of combinations and
directly generate the resulted term.}
As the full calculus of the
suspension notation does, the $\lambdasi$-calculus uses a set
of merging rules to 
combine arbitrarily nested substitutions. Thus the problem with the
sophisticated suspension notation we discussed above also exists
in this context: the complex set of merging rules is not suitable
for real implementations. Requiring the reduction process to
follow head reduction sequences, the merging rules in
$\lambdasi$-calculus can also be simplified to a rule which can
recognize redices in the form of $(\lambdadb t_1)\app t_2$, where
$t_1$ is an explicit substitution term with a certain pattern, and
directly rewrite it into the combination of the substitutions
generated from these two contractions by losing some of the ability
to combine substitutions.

The main difference between the $\lambdasi$-calculus and the
suspension notation is the way they record the adjustment to be
made on indices corresponding to term replacement or renumbering.
In the suspension notation, this adjustment is not explicitly
maintained, but is computed from $nl$ and the natural number
$l$ associated with the environment terms. For example, consider a
suspension term $\env{s,1,nl,(t,l)::nil}$. When the substitution
is to be carried out, the indices of the free variables of $t$
should be increased by $(nl-l)$. In $\lambdasi$-calculus, this
increment information is maintained explicitly within the
environment term as $(t,(nl-l))$. Thus when a substitution is
pushed into an abstraction, this number also needs to be increased
by one, which means all the items in the environment list should
be adjusted. For example, to push a delayed substitution into an
abstraction, $\env{(\lambdadb s),1,nl,(t,l)::nil}$, in the
suspension notation the only work on the environment list is to
add a dummy environment term: $\lambdadb\env{s,2,nl+1,\dum
nl::(t,l)::nil}$. On the other hand, in the $\lambdasi$-calculus,
the already existing environment terms also need to be walked
through to perform the increment: $\lambdadb\env{s,2,nl+1,\dum
1::(t,nl-l+1)::nil}$.

In summary, we believe that the suspension notation provides a
concrete yet sufficiently general basis for examining the use of
explicit substitution systems and the effect of the various
choices afforded by them on actual implementations.

%% file: implicit.tex
\chapter{Environment Based Reduction}\label{chp:implicit}
As we discussed previously, the head normalization process which
reduces the de Bruijn terms under comparison to their head normal
forms when following head reduction sequences is the main issue of the
comparison approach we want to realize for the systems using
$\lambda$-terms to represent syntactic objects. Guided by the
suspension notation, there is still flexibility to choose a
specific strategy to realize the head normalization procedure, and
these choices have different impacts on the heap usage of the
computation systems. Now we discuss these possible
reduction strategies and their impact on heap usage by using
SML procedures for simplicity of exposition and for concreteness,
although the same ideas can be deployed in realizations in any
other language as well. Here we assume a basic familiarity with SML
which one can obtain from \cite{Harper86ECS-LFCS-86-14}. All the
procedures we present are graph-based, $\ie$ $\lambda$-terms are
encoded as directed graphs and destructive changes are used to
register, and thus to share the reduction steps.

\ignore{ In the comparison approach we illustrated in
Chapter~\ref{chp:background}, the reduction process occurs as the
process to reduce the terms under comparison into their head
normal forms following a head reduction sequence, to which we
refer as the head normalization process. Even guided by the
suspension notation, there could be multiple possible choices for
the specific reduction strategies that have different impact on
the heap usage of the computation system using $\lambda$-terms to
represent its syntactic objects. Thus now we discuss the head
normalization procedures guided by the idea of the suspension
notation and using different reduction strategies and their impact
on the heap usage. All the procedures we represent is graph-based,
$\ie$ $\lambda$-terms are encoded as directed graphs and
destructive changes changes are used to register, and thus to
share, the reduction steps.

For simplicity and clarity, all the procedures we present are
encoded in SML. Here we assume a basic familiarity of SML which
one can obtain from \cite{Harper86ECS-LFCS-86-14}. }

\section{An Environment Based Head Normalization Procedure}
According to the suspension notation, it is natural to
consider a reduction procedure based on an environment to achieve the
delaying and combination of substitutions. The most straightforward
way to realize the environment is to use the local
variables and parameters of this reduction procedure. This idea is
encompassed by the first head normalization procedure we present.
In particular, suspensions are realized mainly through the
structure of recursive calls to the normalization routine; they
are not explicitly embedded into terms built on the heap, and thus
the input and output terms of this procedure are pure de Bruijn
terms. In this sense, the suspension notation is used only
implicitly in this reduction strategy.

Figure~\ref{fig:implicitterms} provides the datatype declarations
in SML that serve to represent the structures needed in this
reduction procedure.

\begin{figure}[!ht]
\begin{center}
\begin{verbatim}
datatype rawterm = const of string
 | bv of int
 | fv of string
 | ptr of (rawterm ref)
 | app of (rawterm ref * rawterm ref)
 | lam of (rawterm ref)

type term = (rawterm ref)

datatype eitem = dum of int
 |  bndg of clos * int
and clos = cl of term * int * int * (eitem list)

type env = (eitem list)
\end{verbatim}
\end{center}
\caption{Type declarations for an environment based head
normalization procedure} \label{fig:implicitterms}
\end{figure}

SML expressions of types {\it rawterm} and {\it term} can be
viewed as directed graphs, which are used to support a graph-based
approach to reduction. We refer to such expressions as being
acyclic if the graphs they correspond to in this sense are
acyclic. An important assumption for our later discussion is that
all the SML expressions we deal with are acyclic. In particular,
we expect the input terms of our reduction procedures to hold
this property, and we will show that our reduction procedures
preserve this property.

Among these declarations, the de Bruijn terms are realized as
references to appropriate SML expressions of the type {\it
rawterm}. Correspondingly, the declaration of the type {\it
rawterm} reflects, for the most part, the possible structures of
de Bruijn terms. The constructor {\it ptr} in the declaration of
{\it rawterm} serves to aid the sharing of reduction results which
means that at certain points in our reduction process, we want to
identify (the representations of) terms in a way that makes the
subsequent rewriting of one of them correspond to the rewriting
of the others. Such an identification is usually realized by
representing both expressions as pointers to a common location
whose contents can be changed to effect shared rewritings. In SML
it is possible to update only references and so the common
location itself must be a pointer. The constructor {\it ptr} is
used to encode indirections of this kind when they are needed.

The declaration of the type {\it eitem} reflects the possible
structures of the environment terms. SML expressions of type
{\it clos} are used to record the term paired with an environment,
which are referred to as {\it closures} in the usual leftmost and
outermost reduction control regime. Guided by the suspension
notation, we encode closures in the form of suspensions. This is
the only explicit use of suspensions in this reduction procedure.
The possible appearances of these closures are only at the
top level of the implicit suspensions which are represented by the
explicit terms on the heap together with their environment, $\ie$
they will not be embedded into the structures of other
explicit terms, and will not persist after the termination of the
head normalization procedure.

There are some auxiliary functions that help with manipulation of
the SML expressions of the types in
Figure~\ref{fig:implicitterms}. Under the requirement of
indirections, functions {\it deref} and {\it assign} are used to
look up the value of a term, and to assign one term to another
respectively. Their definitions are given as the following
\begin{verbatim}
fun deref(term as ref(ptr(t))) = deref(t)
  | deref(term) = term

fun assign(t1,ref(ptr(t))) =  assign(t1,t)
  | assign(t1,t2) = t1 := ptr(t2)
\end{verbatim}

Invocations of these two functions on acyclic SML expressions will
obviously terminate and do not introduce cycles if the input
structures are acyclic.

In the course of reduction, we often need to look up a value in
the environment list. The function {\it nth} serves this purpose.
\begin{verbatim}
fun nth(x::l,1) = x
  | nth(x::l,n) = nth(l,n-1)
\end{verbatim}

The environment based head normalization procedure we currently
present essentially has two phases. In the first phase, it traces
a head reduction sequence to produce a head normal form. Once such
a term is exposed, the second phase is entered to compute the
effect of all the substitutions suspended by the first phase.
Procedures {\it hn\_eb} and {\it subst} in
Figure~\ref{fig:hnormimplicit} and Figure~\ref{fig:subst} serve to
implement these two phases respectively. The procedures in
Figure~\ref{fig:construct-functions} are used to update or build
terms on the heap depending on whether the reduction results can
be shared or not.

The procedure {\it hn\_eb} follows head reduction sequences in the
following way. Its last parameter, which has a boolean type, is
used to indicate whether the current term under manipulation is
the function of an application term (when it is set to

\begin{figure}[!ht]
\begin{verbatim}
fun hn_eb(term as ref(bv(i)),0,0,nil,_) = (term,0,0,nil)
 |  hn_eb(term as ref(bv(i)),ol,nl,env,whnf) =
    if (i > ol) then (ref(bv(i-ol+nl)),0,0,nil)
    else (fn dum(l) => (ref(bv(nl - l)),0,0,nil)
           | bndg(cl(t',ol',nl',env'),l) =>
             if (l = nl) then hn_eb(t',ol',nl',env',whnf)
             else hn_eb(t,ol',nl+nl'-l,env',whnf)) (nth(env, i))
 |  hn_eb(term as ref(lam(t)),ol,nl,env,true) = (term,ol,nl,env)
 |  hn_eb(term as ref(lam(t)),ol,nl,env,false) =
    let val (t',ol',nl',env') =
            if (ol = 0) andalso (nl = 0) then hn_eb(t,0,0,nil,false)
            else hn_eb(t,ol+1,nl+1,dum(nl)::env,false)
    in  build_lam(term,t',ol,nl)
    end
 |  hn_eb(term as ref(app(t1,t2)),ol,nl,e,whnf) =
    let val (f,fo,fl,ef) = hn_eb(t1,ol,nl,e,true)
    in (fn ref(lam(t)) =>
           let val s=hn_eb(t,fo+1,fl,bndg(cl(t2,ol,nl,e),nl)::ef,whnf)
           in  update_app(term,s,ol,nl)
           end
         | t => build_app(term,t,t2,ol,nl,e)) (deref(f))
    end
 | hn_eb(ref(ptr(t1)),ol,nl,env,whnf)=hn_eb(deref(t1),ol,nl,env,whnf)
 | hn_eb(term,_,_,_,_)=(term,0,0,nil)
\end{verbatim}
\caption{Head normalization with implicit use of
suspensions}\label{fig:hnormimplicit}
\end{figure}

\begin{figure}[!ht]
\begin{verbatim}
fun subst(ref(app(t1,t2)),ol,nl,env) =
     ref(app(subst(t1,ol,nl,env),subst(t2,ol,nl,env)))
  | subst(ref(lam(t)),ol,nl,env) =
     ref(lam(subst(t,ol+1,nl+1,dum(nl)::env)))
  | subst(ref(bv(i)), ol, nl, env) =
     if i > ol then ref(bv(i+ol-nl))
     else (fn dum(l) => ref(bv(nl - l))
           |  bndg(cl(t,ol',nl',e'),l) =>
                   if (ol'=0) andalso (nl+nl'-l=0) then t
                   else subst(t,ol',nl+nl'-l,e'))(nth(env, i))
  | subst(ref(ptr(t)),ol,nl,env) = subst(deref(t),ol,nl,env)
  | subst(term,_,_,_) = term
\end{verbatim}
\caption{Calculating out suspensions} \label{fig:subst}
\end{figure}

\begin{figure}[!h]
\begin{verbatim}
fun build_lam(term,body,0,0) = (term,0,0,nil)
 |  build_lam(term,body,ol,nl) = (ref(lam(body)),0,0,nil)

fun update_app(term,(t,0,0,nil),0,0) =
    (assign(term,t);
     (t,0,0,nil))
 |  update_app(term,s,ol,nl) = s

fun build_app(term,f,arg,0,0,nil) =
    (assign(term,ref(app(f,arg)));
     (term,0,0,nil))
 |  build_app(term,f,arg,ol,nl,env) =
    (ref(app(f,subst(arg,ol,nl,env))),0,0,nil)
\end{verbatim}
\caption{Construction functions}\label{fig:construct-functions}
\end{figure}
\noindent {\it true}) or not (when it is set to {\it false}).
Consider the case that the input term of {\it hn\_eb} is an
abstraction and its last parameter has the value {\it true}. This
indicates that a head redex, which is required to be contracted
first following head reduction sequences, is exposed. Thus the
recursive call(s) of {\it hn\_eb} returns this abstraction
together with the environment around it. Then {\it hn\_eb}
proceeds to contract this head redex by using rule ($\beta_s$) or
($\beta'_s$), depending on whether the environment around this
abstraction is empty or not. In other words, when the last
parameter of {\it eb\_hn} is set to {\it true}, a weak head normal
form of the incoming implicit suspension is computed, instead of a
head normal form, when it is set to {\it false}. However, at this
time, if the environment around the abstraction is not empty,
$\ie$ the implicit suspension is not trivial, {\it hn\_eb} does
not actually push this implicit suspension into the abstraction as
required by the rewrite rule (r7), and the quadruple returned by
{\it hn\_eb} does not actually represent a weak head normal form
of the incoming implicit suspension, but a ``pre-step" of it. The
reason for this is that implicit suspensions are local to the
reduction procedure, and thus cannot be shared with or interact
with other computation processes. It is unnecessary to explicitly
carry out this propagation, $\ie$ increasing $ol$, $nl$, and
building a dummy environment term, and therefore the effort spent
on it can be saved. For this reason, we are using the ($\beta'_s$)
rule in favor of the following form:
\begin{tabbing}
\qquad\=\kill \> $((\env{\lambdadb t_1,ol,nl,env})\app t_2)
\rightarrow \env{t_1,ol+1,nl,(t_2,nl)::env)}$.
\end{tabbing}
For convenience, we also refer to this ``pre-step'' weak head normal
form as a weak head normal form. Another thing to be noted
here is that, in reality, the returned implicit suspension will be
trivial in all cases other than this one.

Any given term $t$ may be transformed into a head normal form by
invoking the procedure {\it head\_norm1} that is defined as
follows:
\begin{verbatim}
fun head_norm1(t) = hn_eb(t,0,0,nil,false)
\end{verbatim}
At the end of such a call, \emph{t} is intended to be a reference
to a head normal form of its original value as might be expected
in a graph-based reduction scheme. That {\it head\_norm1}
correctly realizes this purpose is the content of the following
theorems.

\input{proof1}

\section{Discussion on Heap Usage}\label{sec:imp-heap}
In the reduction process of this strategy, substitutions
generated by contractions of the head redices of term are delayed
and performed along with the {\it head} reduction steps. In
particular, if the (sub)structures of a term have been {\it
normalized}, then the substitutions involving them are combined
and carried out in one term traversal. Consequently, the new
structures corresponding to such terms are created on the heap
only once. However, in a {\it head} normalization process, once
the head of a head normal form is exposed, the reduction process
terminates without further normalizing of its arguments.
Since all delayed substitutions are maintained locally to
this head normalization procedure, those delayed substitutions to
be performed on the arguments have to be carried out before the
termination of the reduction process and therefore before the
normalization of those arguments. This is not yet the necessary
point at which those substitutions have to be carried out, and
performing substitutions at this point potentially has the
drawback of missing opportunities to combine substitution walks in
the following two situations. First, new redices involving those
arguments could be generated by other computation processes
dynamically, such as the binding of
instantiatable variables after unification which is used in
pattern matching. 
For example, consider a quantified formula such as
$\allx{x}\allx{y} P(x,y)$, where $P(x,y)$ itself represents a
possibly complex formula containing occurrences of $x$ and $y$.
The encoding of this formula using $\lambda$-terms would take the
form $(all\app \lambdax{x} (all\app \lambdax{y}
\overline{P(x,y)}))$, where {\it all} is a constructor chosen to
represent the universal quantifier and $\overline{P(x,y)}$
represents the encoding of $P(x,y)$. In a theorem-proving context
in which a universal quantifier is processed by substitution with
an instantiatable variable, this calculation would be effected by
first recognizing a formula that fits the pattern $(all\app F)$
and then applying the instantiation of $F$ to a new variable. In
particular, when the term $(all\app \lambdax{x} (all\app
\lambdax{y} \overline{P(x,y)}))$ is recognized as fitting the
pattern $(all\app F)$, its subterm $(\lambdax{x} (all\app
\lambdax{y} \overline{P(x,y)}))$ is applied to an instantiatable
variable, say $X$, and hence a head redex is generated. Then the
newly formed application term is head normalized and the head
normal form $(all\app \lambdax{y} \overline{P(X,y)})$ is created.
Note that, using this environment based reduction strategy, the
substitution $[x:=X]$ has already been carried out over
$\overline{P(x,y)}$. After that, the pattern matching process is
invoked on this head normal form again, recognizing that the
incoming term fits the pattern $(all\app F)$, generating a new
application in the form of $(\lambdax{y} \overline{P(X,y)})\app
Y$. The head normalization of this term generates the substitution
$[y:=Y]$ and carries it out over $\overline{P(X,y)}$. Although the
two substitutions $[x:=X]$ and $[y:=Y]$ are performed on the same
term structure $\overline{P(x,y)}$, they are not combined but are
carried out in distinct term traversals. In this situation,
the two redices are generated dynamically and are not revealed to
the same invocation of the head normalization procedure. Thus from
the view of the whole computation process, the performance of the
delayed substitutions over $\overline{P(x,y)}$ each time before
the head normalization procedure terminates is too eager. The
second reason that this reduction strategy misses
opportunities to combine substitution walks over the same term
structures because of the eagerness of the performance of
substitutions, is that there can be redices embedded inside these
arguments on which substitutions are performed, and later
computations could require them to be (head) normalized. For
example, consider the head normalization of the term $((\lambdadb
t_1)\app t_2)$. When this reduction strategy is used, the external
substitution would be percolated over this term, resulting
in a walk over the structure of $t_1$. At a later point, the
embedded redex may be contracted, producing another substitution
traversal over $t_1$. If the substitutions can be delayed until they
are needed, these two distinct walks can actually be combined into
one.

%% file: proof1.tex
\begin{theorem1}\label{th:substcorrect} Let $t$ be a de Bruijn term
and let $\env{t,ol,nl,env}$ be a well-formed suspension. Let {\it
t'} be a reference to the SML expression representing $t$, and
{\it env'} be an SML list representing $env$. Then {\it
subst(t',ol,nl,env')} terminates, preserving the property of
acyclicity and returning a reference to the SML expression
representing a de Bruijn term $r$ that is transformed from
$\env{t,ol,nl,env}$ by applying a series of the reading rules in
Figures~\ref{fig:rewriterules} and~\ref{fig:enhanced-r5}.
\end{theorem1}
\begin{proof}
In the proof of this theorem, we refer to a {\it rewrite sequence}
of a suspension $r_0$ as a sequence $s$ = $r_0$, $r_1$,
$r_2$,...,$r_n$, where $r_n$ is a de Bruijn term; for $i\geq 0$,
there is a suspension term succeeding $r_i$ if $r_i$ is not a de
Bruijn term and, in this case $r_{i+1}$ is obtained from $r_i$ by
the application of one of the reading rules in
Figures~\ref{fig:rewriterules} and~\ref{fig:enhanced-r5}.
Theorem~\ref{th:readingrules-confluent}
and~\ref{th:readingrules-strongnormalizing} assure that every
rewrite sequence of a well-formed suspension terminates at the de
Bruijn term underlying that suspension.

This theorem is proved by induction first on the length of the
longest rewrite sequence of $\env{t,ol,nl,env}$ and then on the
structure of $t'$. Note that in the latter induction, we say that
an SML expression $t'$ is simpler than $s'$ if and only if the
number of value constructors appearing in the structure of $t'$ is
less than that of $s'$. Thus, the latter induction requires our
SML expressions to be acyclic. The preservation of acyclicity
follows easily from the fact that there are no assignments in the
definition of {\it subst}. For the rest, we consider the cases for
the structure of {\it t'}.

The theorem follows obviously if $t'$ is in the form of {\it
ref(const(c))} or {\it ref(fv(f))}.

Suppose that $t'$ is in the form of {\it ptr(s')}. If $s'$ is
acyclic, we know that {\it deref(s')} terminates and returns a
reference to the SML expression representing $t$. Since the
structure of $s'$ is simpler than that of $t'$, by the property of
acyclicity, {\it subst(s',ol,nl,env')} terminates and returns the
SML expression referring to the representation of the de Bruijn
term underlying $\env{t,ol,nl,env}$.

Suppose $t'$ is in the form of {\it ref(bv(i))}. Then the
suspension to be rewritten is $\env{\#i,ol,nl,env}$. There are
three subcases: first, $i>ol$; second, $i\leq ol$ and {\it
nth(env',i)} returns {\it dum(l)}; third, $i\leq ol$ and {\it
nth(env',i)} returns
\begin{tabbing}
\qquad\=\kill \>
{\it bndg(cl($s'_1$,ol',nl',e'),l)}.
\end{tabbing}
The theorem holds  straightforwardly in the first two cases. In the
third case, suppose $s_1$ is the de Bruijn term represented by the
SML expression referred by $s'_1$ and $e$ is the environment
represented by the SML list $e'$. Then we have
\begin{tabbing}
\qquad\=\kill \>
$env[i]=(\env{s_1,ol',nl',e},l)$.
\end{tabbing}
Note that $\env{s_1,ol',nl',e}$ could be a trivial suspension here.
Following reading rule (r11), there is a rewrite step from
$\env{\# i,ol,nl,env}$ to $\env{s_1,ol',nl'+nl-l,e'}$.
Now, if $ol'=0$ and $nl'+nl-l=0$, then
\begin{tabbing}
\qquad\=\kill \>
$\env{s_1,ol',nl'+nl-l,e'}=s_1$, and
\end{tabbing}
$s_1$ is the de Bruijn term underlying the original suspension
$\env{\# i,ol,nl,env}$. If $ol'$ and $nl'+nl-l$ are not both equal
to zero, by the argument already outlined, the length of the
longest rewrite sequence of
\begin{tabbing}
\qquad\=\kill \>
$\env{s_1,ol',nl'+nl-l,e'}$
\end{tabbing}
must be less than that of $\env{\#i,ol,nl,env}$ by at least 1.
By the induction hypothesis,
\begin{tabbing}
\qquad\=\kill \>
{\it subst($s'_1$,ol',nl'+nl-l,e')}
\end{tabbing}
terminates and returns a reference to the SML expression
representing the de Bruijn term $r$ underlying
$\env{s_1,ol',nl'+nl-l,e}$. Further, $r$ is also the de Bruijn
term underlying $\env{\#i,ol,nl,env}$. The theorem follows from
these observations and an inspection of the definition of {\it
subst}.

The cases in which $t'$ is in the form of {\it ref(lam($t'_1$)} and
{\it ref(app($t'_1$,$t'_2$))} both involve the use of a rewrite
rule and hence the proof in these cases invokes the induction
hypothesis based on the length of the longest rewrite sequence of
$\env{t,ol,nl,env}$. We consider in detail the case of {\it
ref(lam($t'_1$))}; the other case is similar.

Suppose that $t'$ is in the form of {\it ref(lam($t'_1$))}. The
suspension to be rewritten is in the form of $\env{\lambdadb
t_1,ol,nl,env}$, where the SML expression representing $t_1$ is
referred to by $t'_1$. Following reading rule (r7), there is a
rewrite step from $\env{\lambdadb t_1,ol,nl,env}$ to
\begin{tabbing}
\qquad\=\kill \>
$\lambdadb \env{t_1,ol+1,nl+1,\dum nl::env}$.
\end{tabbing}
Thus the longest rewrite sequence of suspension
\begin{tabbing}
\qquad\=\kill \>
$\env{t_1,ol+1,nl+1,\dum nl::env}$
\end{tabbing}
is shorter
than that of $\env{\lambdadb t_1,ol,nl,env}$ by at least one. By the
induction hypothesis,
\begin{tabbing}
\qquad\=\kill \>
{\it subst($t'_1$,ol+1,nl+1,dum(nl)::env')}
\end{tabbing}
terminates, and returns a reference to the SML representation of
the term $r$ which is the de Bruijn term underlying
\begin{tabbing}
\qquad\=\kill \> $\env{t_1,ol+1,nl+1,\dum nl::env}$.
\end{tabbing}
Moreover, $(\lambdadb r)$ is the de Bruijn term that $\env{\lambdadb
t_1,ol,nl,env}$ should be rewritten to. From these observations
and an inspection of the code, the theorem follows in this case
too.
\end{proof}

\begin{theorem1}\label{th:implicitcorrect} Let $t'$ be a reference to
the SML expression representing a de Bruijn term $t$ that has a
head normal form. Then {\it head\_norm1(t')} terminates and, when
it does, $t'$ is a reference to the SML expression representing a
head normal form of the original term $t$.
\end{theorem1}
\begin{proof}
Since $t$ has a head normal form, Theorem~\ref{th:HNF} assures
that every (weak) head reduction sequence of $t$ terminates. Hence
we claim the following. If every head reduction sequence (weak
head reduction sequence) of $t$ terminates, then
\begin{tabbing}
\qquad\=\kill \> {\it hn\_eb(t',ol,nl,env',whnf)}
\end{tabbing} terminates,
preserving the acyclicity property and returning a quadruple
\begin{tabbing}
\qquad\=\kill \> {\it (r',rol,rnl,renv')}
\end{tabbing}
representing a head normal form (when {\it whnf} is set to {\it
false}) or a weak head normal form (when {\it whnf} is set to {\it
true}), of $\env{t,ol,nl,env}$, where $env$ is represented by the
SML list $env'$. Further, the returned quadruple is in the form of
{\it (r',0,0,nil)} in all the cases other than that where the 
term that is computed is a weak head normal
form of a non-trivial suspension with an abstraction as its term
skeleton; if $ol=0$, $nl=0$, $env'=nil$ and
$whnf=false$, $t'$ is set to $r'$ at the termination of the
procedure call. The theorem is an immediate consequence of this
claim.

The claim is proved by induction first on the length of the
longest (weak) head reduction sequence of $t$ and then on the
structure of $t'$. By the arguments we mentioned in the previous
theorem, the latter induction requires our SML expressions to be
acyclic. The preservation of acyclicity follows easily from
Theorem~\ref{th:substcorrect} and by observing that the assignments in
functions {\it build\_lam}, {\it update\_app} and {\it build\_app}
won't introduce cycles where these did not exist already. For the
rest, we consider the cases for the structure of {\it t'}.

The claim follows obviously if $t'$ is in the form of {\it
ref(const(c))} or {\it ref(fv(f))}.

Suppose that $t'$ is of the form {\it ref(ptr($s'$))}. If $s'$ is
acyclic, we know that {\it deref(s')} terminates and returns a
reference to the SML expression representing $t$. Since the
structure of $s'$ is simpler than that of $t'$, by the property of
acyclicity,
\begin{tabbing}
\qquad\=\kill \> {\it hn\_eb(s',ol,nl,env',whnf)} \end{tabbing}
terminates and returns a quadruple preserving the properties in
our claim. Thus the claim follows in this case.

Suppose $t'$ is in the form of {\it ref(bv(i))}. If $ol=0$, $nl=0$
and $env'=nil$, the claim holds obviously. Otherwise, the term to
be (weak) head normalized is in fact a non-trivial suspension in
the form of $\env{\#i,ol,nl,env}$. The claim follows obviously in
the case $i>ol$ and the case $i\leq ol$ and {\it nth(env',i)}
returns {\it dum(j)}. Consider the case that $i\leq ol$, and {\it
nth(env',i)} returns {\it bndg(cl(s',ol$1$,nl$1$,env$1$'),l)}. Let $s$
be the de Bruijn term represented by the SML expression referred
by $s'$ and $env1$ be the environment represented by the SML list
$env1'$. Then we have
\begin{tabbing}
\qquad\=\kill \> $env[i]=(\env{s,ol1,nl1,env1},l)$.
\end{tabbing}
Following reading rule (r10), (r11) or (r12), a head reduction
step occurs from the term $\env{\#i,ol,nl,env}$ to
\begin{tabbing}
\qquad\=\kill \> $\env{s,ol1,nl1,env1}$ or
$\env{s,ol1,nl1+nl-l,env1}$.
\end{tabbing}
Hence the the longest (weak) head reduction sequence of  $\env{\#
i,ol,nl,env}$ is longer than that of
\begin{tabbing}
\qquad\=\kill \> $\env{s,ol1,nl1,env1}$ or
$\env{s,ol1,nl1+nl-l,env1}$
\end{tabbing} by at least one.
By the induction hypothesis,
\begin{tabbing}
\qquad\=\kill \> {\it hn\_eb(s',ol1,nl1,env1',whnf)} or {\it
hn\_eb(s',ol1,nl1+nl-l,env1',whnf)}
\end{tabbing}
terminates and returns the quadruple representing a (weak) head
normal form of
\begin{tabbing}
\qquad\=\kill \> $\env{s,ol1,nl1,env1}$ or
$\env{s,ol1,nl1+nl-l,env1}$,
\end{tabbing}
which is also a (weak) head normal form of the term
$\env{\#i,ol,nl,env}$. The claim follows from these observations
and an inspection of the definition of \emph{hn\_eb} in this case.

Suppose {\it t'} has the form of {\it ref(lam(s'))}. Let $s$ be the
de Bruijn term represented by the SML expression referred by $s'$.
Then $t$ is in the form of $\lambdadb s$. Clearly, the
quadruple {\it (t',ol,nl,env')} itself is a weak head normal form
of $\env{\lambdadb s,ol,nl,env}$. Further, if $\env{\lambdadb
s,ol,nl,env}$ is a trivial suspension, {\it (t',ol,nl,env')} is in
the form of {\it (t',0,0,nil)}. Now consider the case that a head
normal form is computed. Suppose
that $ol=0$, $nl=0$ and $env'=nil$. Then
\begin{tabbing}
\qquad\=\kill \> $\env{\lambdadb s,ol,nl,env}=\lambdadb s$.
\end{tabbing}
Since the longest head reduction sequence of $s$ is at most as
long as that of $t$ and the structure of $s'$ is simpler than that
of $t'$, by the induction hypothesis, {\it
hn\_eb(s',0,0,nil,false)} terminates and returns a quadruple in
the form of {\it (r',0,0,nil)} where $r'$ refers to the SML
expression representing a head normal form $r$ of $s$, and
further, $s'$ is updated to a reference to $r'$. It can be seen
that $\lambdadb r$ is a head normal form of $t$, and
correspondingly, $t'$ is a reference to the representation
of $\lambdadb r$ at the point {\it hn\_eb(s',0,0,nil,false)}
terminates. Suppose that $ol$, $nl$ and $env'$ represent a non-empty
environment. There is a head normalization step from the original
suspension $\env{\lambdadb s,ol,nl,env}$ to
\begin{tabbing}
\qquad\=\kill \> $(\lambdadb\env{s,ol+1,nl+1,\dum nl::env})$.
\end{tabbing}
Hence the longest head reduction sequence of
\begin{tabbing}
\qquad\=\kill \> $\env{s,ol+1,nl+1,\dum nl::env}$
\end{tabbing}
is shorter than that of $\env{\lambdadb s,ol,nl,env}$ by at least
one. By the induction hypothesis,
\begin{tabbing}
\qquad\=\kill \> {\it hn\_eb(s',ol+1,nl+1,dum(nl)::env',false)}
\end{tabbing}
terminates and returns a quadruple in the form of {\it
(r',0,0,nil)}, representing a head normal form $r$ of
$\env{s,ol+1,nl+1,\dum nl::env}$. According to reading rule (r7),
$(\lambdadb r)$ is a head normal form of the original suspension.
With these observations and an inspection of the definition of
{\it hn\_eb}, the claim holds in this case.

Suppose {\it t'} is in the form of {\it ref(app($s'_1$,$s'_2$))}.
Let $s^1_1$ and $s^1_2$ be the de Bruijn terms represented by the
SML expressions referred by $s'_1$ and $s'_2$ respectively.

First, consider the case that $ol=0$, $nl=0$ and $env'=nil$. Let
$s^1_1,...,s^k_1,...$ be a weak head reduction sequence of
$s^1_1$, and for $i\leq 1$, $s^{i+1}_2$ is obtained from $s^i_2$
by rewriting some of its subterms that are identical to the weak
head redex of $s^i_1$. Then
\begin{tabbing}
\qquad\=\kill \> $(s^1_1\app s^1_2)$, $(s^2_1\app
s^2_2)$,...$(s^k_1\app s^k_2)$,...
\end{tabbing} is an initial segment of (weak) head reduction
sequence of $t$. Thus, the longest weak head reduction sequence of
$s^1_1$ is at most as long as the longest (weak) head reduction
sequence of $s$. Since the structure of {\it $s'_1$} is simpler
than that of {\it t'}, by the induction hypothesis, {\it
hn\_eb($s'_1$,0,0,nil,true)} terminates and returns the quadruple
which is in the form of {\it ($r'_1$,0,0,nil)} and represents a
weak head normal form of $s^1_1$. Let $r_1$ be the de Bruijn term
represented by the SML expression referred by $r'_1$ and let $r_2$
be the term represented by $s'_2$ at the point {\it
hn\_eb($s'_1$,0,0,nil,true)} terminates. Then there is a (weak)
head reduction step from $t$ to $(r_1\app r_2)$. Now, if $r_1$ is not
in the form of $(\lambdadb x)$, then $(r_1\app r_2)$ is already a
head normal form of the term $t$. Correspondingly, in the definition
of {\it hn\_eb}, {\it t'} is set to refer to the SML expression
representing $(r_1\app r_2)$ via the function {\it build\_app},
and the quadruple to be returned is set to {\it (t',0,0,nil)}.
If
$r_1$ is in the form of $(\lambdadb x)$, then $(r_1\app r_2)$
itself is a (weak) head redex. Following the ($\beta_s$) rule,
there is a head reduction step from the term $(r_1\app r_2)$ to
$\env{r_1,1,0,(r_2,0)::nil}$. Hence, the length of the longest
head reduction sequence of $\env{r_1,1,0,(r_2,0)::nil}$ is as
least smaller than that of $t$ by one. By the induction
hypothesis,
\begin{tabbing}
\qquad\=\kill \>
\emph{hn\_eb($r'_1$,1,0,bndg(cl($s_2$,0,0,nil),0)::nil,whnf)}
\end{tabbing}
terminates and returns a quadruple {\it (r',rol,rnl,renv')} which
represents a (weak) head normal form of
$\env{r_1,1,0,(r_2,0)::nil}$ and is also a (weak) head normal form
of $t$. Further, if $whnf$ is set to {\it false}, this quadruple
must have the form of {\it (r',0,0,nil)}. In the definition of
{\it hn\_eb}, $t'$ is correspondingly set to refer to $r'$ via the
function {\it update\_app}. From these observations and an
inspection of the definition of {\it hn\_eb}, the claim follows.

Now consider the case that $ol$, $nl$ and $env'$ represents a
non-empty environment. The term to be head normalized is in fact
the non-trivial suspension
\begin{tabbing}
\qquad\=\kill \> $\env{(s^1_1\app s^1_2),ol,nl,env}$.
\end{tabbing}
Following the rewriting rules, there is a head reduction step from
that term to
\begin{tabbing}
\qquad\=\kill \> $(\env{s^1_1,ol,nl,env}\app
\env{s^1_2,ol,nl,env})$.
\end{tabbing}
Thus the longest weak head reduction sequence of
$\env{s^1_1,ol,nl,env}$ is shorter than the (weak) head
reduction sequence of
\begin{tabbing}
\qquad\=\kill \> $\env{(s^1_1\app s^1_2),ol,nl,env}$
\end{tabbing}
by at least one. Then by the induction hypothesis,
\emph{hn\_eb($s'_1$,ol,nl,env,true)} terminates and returns a
quadruple {\it ($r'_1$,ol1,nl1,env1')} representing a weak head
normal form of $s^1_1$. Note that if $r'_1$ is not in the form of
{\it ref(lam(x'))}, then $ol1=0$, $nl1=0$ and $env1'=nil$. Let
$\env{r_1,ol1,nl1,env1}$ be this weak head normal form, and $r_2$
be the term represented by what $s'_2$ refers to at this time.
Then there is a (weak) head reduction step from $\env{t,ol,nl,env}$ to
\begin{tabbing}
\qquad\=\kill \>
$(\env{r_1,ol1,nl1,env1}\app\env{r_2,ol,nl,env})$.
\end{tabbing}
Now suppose $r_1$ is not in the form of $\lambdadb x$. Then
$\env{r_1,ol1,nl1,env1}=r_1$. From Theorem~\ref{th:substcorrect},
we know that {\it subst($r_2$,ol,nl,env)} terminates and returns a
reference to a de Bruijn term, say $r_3$, which is the de Bruijn
term underlying $\env{r_2,ol,nl,env}$. Thus the term $(r_1\app r_3)$
is a (weak) head normal form of $\env{t,ol,nl,env}$.
Correspondingly, in the definition of {\it hn\_eb}, the quadruple
to be returned is set to {\it (ref(app($r'_1$,$s'_2$)),0,0,nil)}
via the function {\it build\_app}. On the other hand, if $r_1$ is
in the form of $\lambdadb x$, following rule ($\beta_s'$), there
is a reduction step from term
\begin{tabbing}
\qquad\=\kill \> $(\env{r_1,ol1,nl1,env1}\app\env{r_2,ol,nl,env})$
\end{tabbing}
to term
\begin{tabbing}
\qquad\=\kill \>
$(\env{r_1,ol+1,nl1,(\env{r_2,ol,nl,env},nl1)::env1})$.
\end{tabbing}
Hence the length of the longest head reduction sequence of
\begin{tabbing}
\qquad\=\kill \>
$(\env{r_1,ol+1,nl1,(\env{r_2,ol,nl,env},nl1)::env1})$
\end{tabbing}
is less than that of $\env{t,ol,nl,env}$ by at least one. By the
induction hypothesis,
\begin{tabbing}
\qquad\=\kill \>
\emph{hn\_eb($r'_1$,ol+1,nl1,(cl($r'_2$,ol,nl,env'),0)::env1',whnf)}
\end{tabbing}
terminates and returns the representation of a (weak) head normal
form of
\begin{tabbing}
\qquad\=\kill \>
$\env{r_1,ol+1,nl1,(\env{r_2,ol,nl,env},nl1)::env1}$,
\end{tabbing}
which is also the representation of a (weak) head normal form of
$\env{t,ol,nl,env}$. From these observations and the inspection of
the code, we can see the claim follows in this case.

\end{proof}

%% file: naive.tex
\chapter{Explicit Use of Suspensions}\label{chp:naive}
A way to overcome the potential shortcomings of the reduction
procedure we presented in the previous chapter is to explicitly
build suspensions on the heap. Thus after the termination of one
invocation of the reduction procedure, the suspended substitutions
will persist and therefore can be delayed to the point when it is
actually necessary for them to be carried out. Guided by the
suspension notation, the simplest way to realize such a head
normalization procedure is to build all the suspensions appearing
in the reduction process on the heap. In particular, by matching
the input term structures to the those on the lefthand sides of
rewrite rules, this reduction procedure can choose a proper
rewrite rule to apply, and then explicitly create the term
structures appearing on the righthand side of that rule on the
heap. In this sense, suspensions are used explicitly. In this
chapter, we present a head normalization procedure explicitly
using suspensions in this fashion.

\section{A Head Normalization Procedure with Explicit Suspensions}\label{sec:naive-procedure}

\begin{figure}[!ht]
\begin{verbatim}
datatype rawterm = const of string
  | fv of string
  | bv of int
  | ptr of (rawterm ref)
  | lam of (rawterm ref)
  | app of (rawterm ref) * (rawterm ref)
  | susp of (rawterm ref)*int*int*(eitem list)
and eitem = dum of int
  | bndg of (rawterm ref) * int

type env = (eitem list)

type term = (rawterm ref)
\end{verbatim}
\caption{Type declarations for suspension terms}
\label{fig:suspterms}
\end{figure}

The datatype declarations used in this head normalization
procedure are presented in Figure~\ref{fig:suspterms}. They differ
from those in Figure~\ref{fig:implicitterms} in the following two
aspects: first, suspensions are explicitly accepted as a possible
term structure of the type {\it rawterm} and denoted by the
constructor {\it susp}; second, the structures appearing in the
environment can be an arbitrary term as opposed to only closures
in the environment based head normalization procedure in
Chapter~\ref{chp:implicit}.

The auxiliary functions \emph{deref}, \emph{assign}, and
\emph{nth} are still available to this procedure.

\begin{figure}[!t]
\begin{verbatim}
fun lazy_read(term as ref(susp(t,ol,nl,env))) =
    lazy_read_aux(term,deref(t),ol,nl,env)
  | lazy_read(_) = ()
and lazy_read_aux(t1,ref(bv(i)),ol,nl,env) =
    if i > ol then t1 := bv(i + nl - ol)
    else ((fn dum(l) => t1 := bv(nl - l)
           |  bndg(t2,l) =>
              (if (nl = l) then assign(t1,t2)
               else ((fn ref(susp(t3,ol',nl',e')) =>
                         t1 := susp(t3,ol',nl'+ nl - l,e')
                      | t => t1 := susp(t,0,nl - l,nil)
                   ) (deref t2));
               (lazy_read t1))) (nth (env,i)))
  | lazy_read_aux(t1,ref(app(t2,t3)),ol,nl,env) =
    t1 := app(ref(susp(t2,ol,nl,env)),ref(susp(t3,ol,nl,env)))
  | lazy_read_aux(t1,ref(lam(t2)),ol,nl,env) =
    t1 := lam(ref(susp(t2,ol+1,nl+1,dum(nl)::env)))
  | lazy_read_aux(t1,t,ol,nl,env) =
    (lazy_read(t) ; lazy_read_aux(t1,deref(t),ol,nl,env))
  | lazy_read_aux(t1,t2,_,_,_) = t1 := !t2
\end{verbatim}
\caption{Auxiliary procedure for exposing term structures under
suspensions} \label{fig:naivered1}
\end{figure}

\begin{figure} [!ht]
\begin{verbatim}
fun beta_contract(term,t1 as ref(susp(t3,ol,nl,dum(nl1)::e)),t2)=
    if nl = nl1+1 then term := susp(t3,ol,nl1, bndg(t2,nl1)::e)
    else term := susp(t1,1,0,[bndg(t2,0)])
  | beta_contract(term,t1,t2) = term := susp(t1,1,0,[bndg(t2,0)])

fun hn_ex(term as ref(app(t1,t2)),whnf) =
    (hn_ex(t1,true) ;
     (fn ref(lam(t)) => (beta_contract(term,t,t2);
                         hn_ex(term,whnf))
       | _  => ()) (deref t1))
  | hn_ex(ref(lam(t)),false) = hn_ex(t,false)
  | hn_ex(term as ref(susp(_,_,_,_)),whnf) =
        (lazy_read(term) ; hn_ex(term,whnf))
  | hn_ex(term as ref(ptr(t)),whnf) =
      (hn_ex(t,whnf) ; assign(term,t))
  | hn_ex(_,_) = ()
\end{verbatim}
\caption{Head normalization using suspensions and immediate
rewriting} \label{fig:naivered2}
\end{figure}

This reduction approach involves the creation of
representations on the heap for all the new structures that appear on the
righthand side of a rule immediately on the application of that
rule. Thus, suppose that at a certain point in computation, we use
the rule
\begin{tabbing}
\quad\=\kill \> $\env{(t_1\app t_2),ol,nl,e} \ra
(\env{t_1,ol,nl,e}\app \env{t_2,ol,nl,e})$
\end{tabbing}
\noindent for propagating substitutions over applications. In the
mode that we are presently considering, we will create the new
structures $\env{t_1,ol,nl,e}$ and $\env{t_2,ol,nl,e}$ and
destructively update the term on the lefthand side with an
application formed out of these two pieces before proceeding to
the next step in reduction.\ignore{In this reduction procedure,
our purpose is to use the rewrite rules of the suspension notation
to conduct the reduction process. Since all the substitution
information are already embedded in the input term structure,
purely depending on the structure of the incoming term, the
procedure can determine which rewrite rule to apply and after the
application, the resulted term structures on the right hand side
of the rule are created on the heap explicitly.} Thus once a head
normal form is exposed, this procedure has no problem in terminating
immediately without accessory operations to carry out the
suspended substitutions. A consequence of this approach is that it
should include mechanisms for incrementally `unravelling' the
suspensions met during the reduction processing.

Procedure {\it lazy\_read} in Figure~\ref{fig:naivered1} is used
to incrementally expose the top-level non-suspension structure
from a suspension when such a term is met during the normalizing
process. Procedure {\it beta\_contract} in
Figure~\ref{fig:naivered2} serves to determine which of the
($\beta_s$) and ($\beta'_s$) rules is appropriate to use when a
$\beta$-redex has been discovered and to effect the corresponding
rewriting step. Procedure {\it hn\_ex} in
Figure~\ref{fig:naivered2} realizes the overall control of the
reduction process. Similar to the procedure {\it hn\_eb}, {\it hn\_ex} can
also be invoked in two modes according to its last parameter with
{\it true} to generate a weak head normal form and {\it false} to
generate a head normal form, and follows the head reduction
sequences in the same way that {\it hn\_ex} does. Note that in this
procedure, first, all the term structures are created
explicitly on the heap and thus can be shared by other computation
processes, and second, once a suspension is met, the non-suspension
structure resulting from the application of one of the reading
rules must be created on the heap for the reduction procedure to
progress; therefore, the propagation of a suspension in the form of
$\env{\lambdadb t,ol,nl,env}$ over the abstraction inside can not
be avoided. Thus the weak head normal forms of suspensions with
abstractions as their term skeletons in this reduction procedure
are their actual weak head normal forms in the form of
$\lambdadb\env{t,ol+1,nl+1,\dum nl::env}$, as opposed to the
``pre-steps'' of them in the form of $\env{\lambdadb
t,ol,nl,env}$.

Any given term $t$ may be transformed into a head normal form by
invoking the procedure {\it head\_norm2} that is defined as
follows:
\begin{verbatim}
fun head_norm2(t) = hn_ex(t,false)
\end{verbatim}
At the end of such a call, {\it t} is intended to be a reference
to a head normal form of its original value as might be expected
in a graph-based reduction scheme. That {\it head\_norm2}
correctly realizes this purpose is the content of the following
theorem.

\begin{theorem1}\label{th:naivecorrect} Let $t$ be a reference to
the representation of a suspension term that translates via the
reading rules to a de Bruijn term with a head normal form. Then
{\it head\_norm2(t)} terminates and, when it does, $t$ is a
reference to the representation of a generalized head normal form
of the original term.
\end{theorem1}

\begin{proof}
See \cite{nadathur99finegrained}.
\end{proof}

\section{Discussion on Heap Usage}
With the ability to record suspensions on the heap, this head
normalization procedure need not carry out the delayed
substitutions on the arguments of a head normal form before its
termination. Consequently, this head normalization procedure has
the ability to combine the substitutions caused by contractions of
redices dynamically generated across computation steps and by
contractions of redices nested inside the arguments. However,
there is still a significant drawback in its heap usage: in the course of
reduction, many term structures resulting from the application of
the rewrite rules are only intermediate to the reduction process,
because once created, they are immediately rewritten by the next
application of a rewrite rule. In this sense, these kinds of terms
are in fact only local to the reduction procedure. The creation of
such local terms on the heap is certainly unnecessary. For
example, consider the rule for propagating substitutions over
applications:
\begin{tabbing}
\quad\=\kill \> $\env{(t_1\app t_2),ol,nl,e} \ra
(\env{t_1,ol,nl,e}\app \env{t_2,ol,nl,e})$
\end{tabbing}
\noindent An eager creation of the structures $\env{t_1,ol,nl,e}$,
$\env{t_2,ol,nl,e}$ and the application on the righthand side has
the potential for using heap space unnecessarily: the very next
steps may require the first of these suspensions to be rewritten
and, a few steps later, the outer application itself may be
recognized as a $\beta$-redex. This problem is avoided by the
reduction procedure we describe in Chapter~\ref{chp:implicit},
because it utilizes the recursion stack to record the intermediate
terms and only commits structures on the heap when these are known
to be necessary.

\ignore{ The purpose of reading rules of the suspension notation
is to direct how to push a suspension over a certain term
structure. Depending on the term structure, there could be many
such suspension propagation steps before a head normal form is
exposed. In most cases the intermediate term structures will be
abandoned immediately by the next rewriting step. For example, a
suspension has the form of $\env{(\lambdadb t_1)\app t_2,
ol,nl,env}$ is to be reduced. In this reduction process, two new
suspension terms $\env{\lambdadb t_1,ol,nl,env}$ and
$\env{t_2,ol,nl,env}$ will be explicitly built on the heap and the
original suspension term will be destructively updated to the
application of these two terms according to rule (r6). Then rule
(r7) is applied to the function $\env{\lambdadb t_1,ol,nl,env}$ to
build a new term $\lambdadb\env{t_1,ol+1,nl+1,\dum nl::env}$ and
term $\env{\lambdadb t_1,ol,nl,env}$ is abandoned. Now a redex
appears on the top level of this term and should be reduced. After
this reduction, the structure $\lambdadb\env{t_1,ol+1,nl+1,\dum
nl::env}$ built in the previous step is also abandoned. In the
environment based head reduction procedures, since most of the
intermediate suspensions are recorded in recursion stacks, such
kind of heap usage is avoided.}

%% file: combined.tex
\chapter{Combining Implicit and Explicit Uses of
Suspensions}\label{chp:combined} The two head reduction procedures
we presented in the previous chapters have complementary benefits
and drawbacks. The environment based head normalization procedure
in Chapter~\ref{chp:implicit} effectively utilizes local
variables and parameters to avoid the creation of intermediate
term structures in the head reduction process, but fails to delay
the substitutions to be performed on the arguments of the head
normal forms out of one invocation of the reduction procedure.
The procedure in Chapter~\ref{chp:naive} delays the
substitutions to be performed on the arguments of the head normal
forms out of one invocation of the reduction procedure in the form
of explicit suspensions on the heap, but builds all the
intermediate term structures of the head reduction process on the
heap too. Now we present a synthesis of those two reduction
procedures, and then combine the benefits of both. The essential
idea is to follow the basic regime of the environment based
reduction process in Chapter~\ref{chp:implicit}, but once a head
normal form is found, to explicitly build suspensions on the heap
to delay the substitutions further out of the current invocation
of the reduction procedure.


\begin{figure}[t]
\begin{verbatim}
fun build_lam(term,body,0,0) = (term,0,0,nil)
 |  build_lam(term,body,ol,nl) = (ref(lam(body)),0,0,nil)

fun update_app(term,(t,0,0,nil),0,0) =
    (assign(term,t);
     (t,0,0,nil))
 |  update_app(term,s,ol,nl) = s

fun build_app(term,f,arg,0,0,nil) =
    (assign(term,ref(app(f,arg)));
     (term,0,0,nil))
 |  build_app(term,f,arg,ol,nl,env) =
    (ref(app(f,ref(susp(arg,ol,nl,env)))),0,0,nil)

fun mk_explicit(term,(t,0,0,nil),_,_) =
    (assign(term,t); (t,0,0,nil))
 |  mk_explicit(term,(t,ol,nl,env),0,0) =
    (assign(term, ref(susp(t,ol,nl,env))); (t,ol,nl,env))
 |  mk_explicit(term,(ref(lam(t)),ol,nl,env),ol',nl') =
    (assign(term, ref(lam(ref(susp(t,ol+1,nl+1,dum(nl)::env)))));
     (term,0,0,nil))

fun arg(t,0,0,nil) = t
 |  arg(t,ol,nl,env) = (ref(susp(t,ol,nl,env)))
\end{verbatim}
\caption{Construction functions}\label{fig:construct-functions2}
\end{figure}

In order to achieve the ability to explicitly build suspensions
over arguments of the head normal forms, it is necessary to use
the richer representation of terms that includes an encoding of
suspensions. Assuming the datatype declarations in
Figure~\ref{fig:suspterms} and the accessory function {\it deref},
{\it assign}, and {\it nth}, a collection of SML
\begin{figure}[!ht]
\begin{verbatim}
fun hn_co(term as ref(bv(i)),0,0,[],_) = (term,0,0,[])
 |  hn_co(term as ref(bv(i)),ol,nl,env,w) =
    if (i > ol) then (ref(bv(i+ol-nl)),0,0,nil)
    else (fn dum(l) => (ref(bv(nl-l)),0,0,nil)
           | bndg(t,l) =>  if (nl = l) then hn_co(t,0,0,nil,w)
             else (fn ref(susp(t2,o,n,e))=>hn_co(t2,o,n+nl-l,e,w)
                    | t=>hn_co(t,0,nl-l,[],w))(deref(t)))(nth(env,i))
 |  hn_co(term as ref(lam(t)),ol,nl,env,true) = (term,ol,nl,env)
 |  hn_co(term as ref(lam(t)),ol,nl,env,w) =
    let val (t',o,n,e)=if (ol=0) andalso (nl=0) then hn_co(t,0,0,[],w)
                       else hn_co(t,ol+1,nl+1,dum(nl)::env,w)
    in build_lam(term,t',ol,nl) end
 |  hn_co(term as ref(app(t1,t2)),ol,nl,env,w) =
    let val (f,fo,fl,fe) = hn_co(t1,ol,nl,env,true)
    in (fn ref(lam(t))=>
           let val s=hn_co(t,fo+1,fl,bndg(arg(t2,ol,nl,env),nl)::fe,w)
           in  update_app(term,s,ol,nl) end
         | t => build_app(term,t,t2,ol,nl,env))(deref(f)) end
 |  hn_co(term as ref(susp(t,ol,nl,env)),ol',nl',env',whnf) =
    let val s = mk_explicit(term,hn_co(t,ol,nl,env,whnf),ol',nl')
    in  if (ol'=0) andalso (nl'=0) then s
        else hn_co(term,ol',nl',env') end
 |  hn_co(ref(ptr(t)),ol,nl,env,whnf) = hn_co(deref(t),ol,nl,env,whnf)
 |  hn_co(term,_,_,_,_) = (term,0,0,nil)
\end{verbatim}
\caption{Head normalization using suspensions implicitly and
explicitly} \label{fig:hnormcombined}
\end{figure}

\noindent functions that utilize the proposed idea is presented
in Figures~\ref{fig:construct-functions2}
and~\ref{fig:hnormcombined}.

Among these functions, {\it hn\_co} actually performs the main
work in the reduction. Functions {\it build\_lam}, {\it
update\_app} and {\it build\_app} serve to update or create new
term structures on the heap according to whether the reduction
results can be shared or not. It can be observed that these
functions are in most respects identical to the environment based
procedures we presented in Chapter~\ref{chp:implicit}. However,
there are two significant differences. First, once a non-reducible
head of an application is exposed, as opposed to carrying out the
delayed substitutions on its argument as {\it hn\_eb} does by
calling {\it subst}, this reduction procedure directly builds a
non-trivial suspension over the argument and constructs a new
application term having this suspension as its argument. Second,
suspensions should also be considered as a possible term category
the reduction procedure could encounter. It is interesting to note
that, in this case, if the incoming suspension is under a
non-empty environment, in order to preserve the ability to commit
structures to the heap only when necessary, the embedded
suspension needs to be processed first. This
reduction order is different from the one used by the reduction
procedure in Chapter~\ref{chp:naive}, which commits structures to
the heap eagerly, and actually is not leftmost and outermost.
However, the progression of reduction steps is still encompassed
by the notion of a (weak) head reduction sequence in
Definition~\ref{def:susp-hrs}. Function {\it mk\_explicit} serves
to update the explicit suspension to its (weak) head normal form.
Consider the situation in which a weak head normal form of an explicit
suspension is computed, and this weak head normal form has an
abstraction as its term skeleton. For the reason we discussed in
Chapter~\ref{chp:implicit}, the quadruple returned by {\it hn\_co}
does not actually represent a weak head normal form of this
suspension in the form of $\lambdadb\env{t,ol+1,nl+1,\dum
nl::env}$, but a ``pre-step'' of it in the form of $\env{\lambdadb
t,ol,nl,env}$. However, if the environment around this suspension
is not empty, $\ie$, the explicit suspension is embedded in an
implicit one, in order to make the reduction procedure progress,
we have to actually propagate the suspension into the abstraction,
and update the explicit suspension to its actual weak head normal
form: $\lambdadb\env{t,ol+1,nl+1,\dum nl::env}$. This situation is
also taken cared of by the function {\it mk\_explicit}.

\ignore{In order to explicitly build suspensions on the heap,  the
data types of this procedure needs to accept suspensions as a
possible input term category. For this reason, the data type
declarations of this procedure are the same as that of {\it
hnorm2} as illustrated in Figure~\ref{fig:suspterms}.
Figure~\ref{fig:hnormcombined} presents the definition of this
reduction procedure {\it hnorm3} and the auxiliary function {\it
make\_explicit}. The recursion control of this procedure mainly
follows that of {\it hn\_eb}. However, there are two significant
differences. First, once a non-reducible head of an application is
recovered, opposed to carrying out the delayed substitutions on
the argument of the application as {\it hnorm1} does, this
reduction procedure will directly build a non-trivial suspension
term into the argument part of the application. Second,
suspensions should also be considered as a possible term category
the reduction procedure could encounter. It is interesting to note
that, if the incoming suspension term is under a non-empty
environment, in order to preserve the ability to commit structures
to heap only when necessary, the embedded suspension needs to be
processed first in this case. This order is different from the one
used by the reduction procedure of the previous chapter that
commits structures to heap eagerly, which is not the leftmost,
outmost order. However, the progression of reduction steps is
still encompassed by the generalized notion of head reduction
sequence in Definition~\ref{def:HRS}.}

Any given term $t$ may be transformed into head normal form by
invoking the procedure {\it head\_norm3} that is defined as
follows:
\begin{verbatim}
fun head_norm3(t) = hn_co(t,0,0,nil,false)
\end{verbatim}
At the end of such a call, \emph{t} is intended to be a reference
to a head normal form of its original value as might be expected
in a graph-based reduction scheme. The correctness of {\it
head\_norm3} is the content of the following theorem whose proof
is similar to that of Theorem~\ref{th:implicitcorrect}.

\input{proof2}

%% file: proof2.tex
\begin{theorem2} Let {\it t'} be a reference to
the representation of a suspension term $t$ that translates via
the reading rules to a de Bruijn term with a head normal form.
Then {\it head\_norm3(t')} terminates and, when it does, $t'$ is a
reference to the representation of a generalized head normal form
of the original term.
\end{theorem2}
\begin{proof}
Since $t$ has a head normal form, Theorem~\ref{th:HNF} assures
that every (weak) head reduction sequence of $t$ terminates. As in
the proof of Theorem~\ref{th:implicitcorrect}, we therefore claim
the following. If every head reduction sequence (weak head
reduction sequence) of $t$ terminates, then
\begin{tabbing}
\qquad\=\kill \> {\it hn\_co(t',ol,nl,env',whnf)}
\end{tabbing} terminates,
preserving the acyclicity property and returning a quadruple
\begin{tabbing}
\qquad\=\kill \> {\it (r',rol,rnl,renv')}
\end{tabbing}
representing a head normal form (when {\it whnf} is set to {\it
false}) or a weak head normal form (when {\it whnf} is set to {\it
true}) of $\env{t,ol,nl,env}$, where $env$ is represented by the
SML list $env'$. Further, the returned quadruple is in the form of
{\it (r',0,0,nil)} in all the cases other than that where 
a weak head normal
form of $\env{t,ol,nl,env}$ is computed and this weak head normal
form is a non-trivial suspension with an abstraction as its term
skeleton; if $ol=0$, $nl=0$, $env'=nil$ and $whnf=false$, $t'$ is
set to $r'$ at the termination of the procedure call. The theorem
is an immediate consequence of this claim.

The claim is proved by induction first on the length of the
longest (weak) head reduction sequence of $t$ and then on the
structure of {\it t'}. The preservation of acyclicity follows
easily by observing that the assignments in {\it hn\_co},
{\it make\_explicit}, {\it build\_app}, {\it build\_lam}
and {\it update\_app} do not introduce cycles if these did not
exist already. For the rest, we consider the cases for the
structure of \emph{t'}.

When {\it t'} is in the form of {\it ref(const(c))}, {\it
ref(fv(c))}, {\it ref(ptr($t'_1$))}, {\it ref(bv(i))}, or {\it
ref(lam(s'))}, the proof of this claim is exactly the same as that
of Theorem~\ref{th:implicitcorrect} in such cases.

Now suppose {\it t'} is in the form of {\it
ref(app($s'_1$,$s'_2$))}. The proof of this claim is the same as
that of Theorem~\ref{th:implicitcorrect} in all the cases other
than the following: the environment represented by $ol$, $nl$ and
$env'$ is not empty, and {\it hn\_co($s'_1$,ol,nl,env',true)} returns
the quadruple {\it ($r'_1$,ol$1$,nl$1$,env$1$')},
 where $r'_1$ is not in the form of
{\it ref(lam(x'))}. Note that in this case, $ol1=0$, $nl1=0$ and
$env1'=nil$ and a head of the (weak) head normal form of
$\env{t,ol,nl,env}$ is exposed. Let $r_1$ be the term represented
by the SML expression referred to by $r'_1$ and let $r_2$ be the term
represented by the SML expression referred to by $s'_2$ at the point
that {\it hn\_co($s'_1$,ol,nl,env',true)} terminates. Then the
term $(r_1\app\env{r_2,ol,nl,env})$ is a (weak) head normal
form of $\env{t,ol,nl,env}$. Correspondingly, in the definition of
{\it hn\_co}, the quadruple to be returned is set to
\begin{tabbing}
\qquad\=\kill \>
{\it(ref(app($r'_1$,ref(susp($s'_2$,ol,nl,env)))),0,0,nil)}
\end{tabbing}
via the function {\it build\_app}.
Therefore, the claim holds in this case.

Now consider the case that {\it t'} is in the form of
{\it ref(susp(s',ol1,nl1,env1'))}
 which is a reference to the
SML expression representing the term $\env{s,ol1,nl1,env1}$. Since
the structure of the term {\it s'} is simpler than that of {\it
ref(susp(s',ol1,nl1,env1'))}, by the induction hypothesis, {\it
hn\_co(s',ol1,nl1,env1,whnf)} terminates and returns a quadruple
{\it (r',rol,rnl,renv')} representing a (weak) head normal form of
$\env{s,ol1,nl1,env1}$.

If {\it ol=0, nl=0, env'=nil}, then $\env{t,ol,nl,env}=t$, and
clearly the suspension represented by {\it (r',rol,rnl,renv')} is
already a (weak) head normal form of $t$. Further, if {\it whnf}
is {\it false}, by the induction hypothesis, {\it
(r',rol,rnl,renv')} must have the form of {\it (r',0,0,nil)}.
Correspondingly, in the definition of {\it hn\_co}, via the
function {\it mk\_explicit}, $t'$ is updated to a reference to
$r'$ if {\it (r',rol,rnl,renv')} is in the form of {\it
(r',0,0,nil)}, and is updated to a reference to {\it
ref(susp(r',rol,rnl,env'))} otherwise.

On the other hand, if the environment represented by {\it ol},
{\it nl} and {\it env'} is not empty, the term to be (weak) head
normalized is in fact in the form of
\begin{tabbing}
\qquad\=\kill \> $\env{\env{s,ol1,nl1,env1},ol,nl,env}$.
\end{tabbing}
Suppose $r$ is the term represented by the SML expression
referred to by $r'$ and $renv$ is the environment represented by the
SML list $renv'$.
Then the (weak) head reduction sequence of
\begin{tabbing}
\qquad\=\kill \> $\env{\env{s,ol1,nl1,env1},ol,nl,env}$
\end{tabbing}
must have the form
\begin{tabbing}
\qquad\=\kill \> $\env{\env{s,ol1,nl1,env1},ol,nl,env}$, ...,
$\env{\env{r,rol,rnl,renv},ol,nl,env}$,....
\end{tabbing}
Now if $r$ is in the form of $(\lambdadb x)$, and $\env{r,rol,rnl,renv}$
is not a trivial suspension, then the term
\begin{tabbing}
\qquad\=\kill \>$\env{\lambdadb\env{x,rol+1,rnl+1,\dum
rnl::renv},ol,nl,env}$
\end{tabbing}
must occur somewhere in the reduction sequence of
\begin{tabbing}
\qquad\=\kill \>$\env{\env{r,rol,rnl,renv},ol,nl,env}$,
\end{tabbing}
and therefore in the reduction sequence of
\begin{tabbing}
\qquad\=\kill \> $\env{\env{s,ol1,nl1,env1},ol,nl,env}$.
\end{tabbing}
Thus the longest (weak) head reduction sequence of
\begin{tabbing}
\qquad\=\kill \> $\env{\env{s,ol1,nl1,env1},ol,nl,env}$
\end{tabbing}
is longer than that of
\begin{tabbing}
\qquad\=\kill \>$\env{\lambdadb\env{x,rol+1,rnl+1,\dum
rnl::renv},ol,nl,env}$.
\end{tabbing}
Let $x'$ be the reference to the SML expression representing $x$.
By the induction hypothesis,
\begin{tabbing}
\qquad\=\kill \>
{\it hn\_co(ref(lam(ref(susp(x',rol+1,rnl+1,dum(rnl)::renv')))),ol,nl,env',whnf)}
\end{tabbing}
terminates and returns a quadruple representing a (weak) head
normal form of
\begin{tabbing}
\qquad\=\kill \> $\env{\env{s,ol1,nl1,env1},ol,nl,env}$.
\end{tabbing}
Now suppose $\env{r,rol,rnl,renv}$ is trivial (note that if $r$ is
not an abstraction, then this suspension must be a trivial one by
our induction hypothesis), then
\begin{tabbing}
\qquad\=\kill \>$\env{\env{r,rol,rnl,renv},ol,nl,env}$
\end{tabbing}
is in fact $\env{r,ol,nl,env}$.
Thus the longest (weak) head reduction sequence of
\begin{tabbing}
\qquad\=\kill \> $\env{\env{s,ol1,nl1,env1},ol,nl,env}$
\end{tabbing}
is longer than that of
\begin{tabbing}
\qquad\=\kill \>$\env{r,ol,nl,env}$.
\end{tabbing}
By the induction hypothesis,
\begin{tabbing}
\qquad\=\kill \>
{\it hn\_co(r',ol,nl,env',whnf)}
\end{tabbing}
terminates and returns a quadruple representing a (weak) head
normal form of
\begin{tabbing}
\qquad\=\kill \> $\env{\env{s,ol1,nl1,env1},ol,nl,env}$.
\end{tabbing}
The claim follows from these observations and an inspection of the
definition of {\it hn\_co}.
\end{proof}

%% file: comparison1.tex
\chapter{Comparisons of Different Head Reduction Strategies}\label{chp:comparison}
In this chapter, we consider a quantification of the relevance in
practice of the intuitions underlying the various reduction
procedures discussed in the earlier chapters.

Our experiment is based on the higher-order logic programming
language $\lambda$Prolog. This language employs $\lambda$-terms as
a means for realizing higher-order approaches to the processing of
syntactic structure. Thus, within it, $\lambda$-terms are
available for use in representing objects whose understanding
embodies binding notions, and operations such as higher-order
unification and reduction can be utilized for manipulating such
representations in logically meaningful ways. By running a variety
of actual $\lambda$Prolog programs and collecting suitable data
over these, we can therefore obtain an understanding of the impact
of the different approaches to reduction. At the computation
level, the use $\lambda$Prolog  makes of $\lambda$-terms is quite
similar to what is done in logical frameworks, proof assistants
and metalanguages such as Twelf, Isbelle and Coq. The observations
we make relative to this language therefore carry over naturally
to all these other contexts.

We have carried out the described idea by taking advantage of a
compiler and abstract machine based implementation of
$\lambda$Prolog called \emph{Teyjus}. This system, implemented in
the C language, supports a low-level encoding of $\lambda$-terms
based on the suspension notation. Reduction computations within it
are isolated in a head normalization procedure. Thus we can easily
vary the reduction strategies used in this procedure and measure
the effects of these variations. As a basis for our study, we have
implemented three different head normalization
procedures following the lines of discussion in this thesis, and
we have metered these to collect information about the number of
heap cells created over the entire duration of any given user
program.

\section{Experiment Examples}
The data that we provide have been obtained by running the
following representative user programs:
\begin{itemize}
\item {\it [Compiler]} \ This is a compiler for a small imperative language
with object-oriented features. 
This program includes a bottom-up parser, a continuation
passing-style intermediate language, and generation of native byte
code. Significant parts of the computation in this program do not
in fact involve $\lambda$-terms. However, there are also major
parts that do and our study reveals that choices in reduction
strategies here can have a significant impact on behavior.

\item {\it [Typeinf]} \ This is a program that infers principal type schemes
for ML-like programs. The representation of types
treats quantification explicitly within this program and
abstraction in the metalanguage is used to capture the binding
effect. 
Given the treatment of type variables, unification over
types is explicitly programmed. Thus, many of the typical features
of a metalanguage are exercised by this program.

\item {\it [Hilbert]} \ This is an encoding in $\lambda$Prolog of the
process of solving diophantine equations through higher-order
unification. 
Solutions are not generated
completely by this program in many instances. Rather, solvability
is often determined, the exact identity of solutions being
dependent on the unifiers for `flexible-flexible' disagreement
pairs left behind at the end of the computation.

\item {\it [Funtrans]} \  This is an collection of transformations on
functional programs, such as through partial
evaluation.

\item {\it [SKI]} \ This program realizes an object-level head normalization
on arbitrary compositions of the well-known combinators {\it S},
{\it K} and {\it I}. The data that is collected is based on the
application of this procedure to a collection of five hundred
combinator compositions that were created with help from a random
number generator.

\item {\it [Church]} \ This program involves arithmetic
calculations based on Church's encoding of numerals and the
combinators for addition and multiplication.  The largest `number'
used in this program is around twenty thousand.
\end{itemize}

The first two programs exemplify what might be called the
L$_\lambda$ style of programming \cite{Miller91jlc}. As an
programming idiom, this is a popular one amongst $\lambda$Prolog,
Elf and Isabelle users and, in fact, arguably the most important
case to consider in performance assessments. Computations in this
class proceed by first dispensing with all abstractions in
$\lambda$-terms using new constants, then carrying out a
first-order style analysis over the remaining structures and
eventually abstracting out the new constants. The process of
abstraction elimination is realized in the following way:
once an abstraction is recognized by the pattern matching process,
it is applied to a new constant and thus a redex is generated.
After that, the head normalization process is invoked on the newly
formed application. In this sense, the generation of redices is
interleaved with the head reduction process and thus most redices
are not revealed to one invocation of the head normalization procedure.
The unification operation in such a language subset is known to be
deterministic, and most of the redices have the characteristic that
the arguments of them are all constants.

Programs {\it Hilbert} and {\it Funtrans} include cases of genuine
higher-order unification calculations. Unlike the unification
process in the L$_\lambda$ class that always terminates and
returns the unique unifier, there may be branching in unification.
In particular, in these cases,
different $\lambda$-terms may have to be posited as
bindings for instantiatable variables and reductions and other
computations would have to be carried out, and possibly
backtracked over, using such terms.

Programs {\it SKI} and {\it Church} represent a situation in which
$\lambda$-terms are used mainly in reduction, the unification
computation is largely first-order in nature.

\section{Experiment Results}
\begin{figure}[!ht]
\begin{center}
\begin{tabular}{|l|l|l|l|} \hline
 &implicit suspensions     &explicit suspensions    &combination approach\\\hline
{\it [Compiler]}
    &2,640,909             &635,851                &134,316 \\ \cline{2-4}
    &19,703,580            &6,372,836              &1,764,172 \\ \hline
{\it [Typeinf]}
    &7,142,880            &7,691,000             &1,722,696\\ \cline{2-4}
    &61,045,064           &80,893,824            &22,012,896   \\ \hline
{\it [Hilbert]}
   &170,123                &18,894                 &5,642\\ \cline{2-4}
   &1,387,192              &196,596                &75,020   \\ \hline
{\it [Funtrans]}
   &28,027                 &62,277                 &24,404\\ \cline{2-4}
   &319,740                &620,632                &313,744  \\ \hline
{\it [SKI]}
   &98,319                 &180,777               &76,779\\ \cline{2-4}
   &1,164,656              &1,800,160             &939,104  \\ \hline
{\it [church]}
   &44,797                 &137,936                 &37,162\\ \cline{2-4}
   &610,528                &1,342,076               &531,892\\ \hline

\end{tabular}
\end{center}
\caption{Heap usage for different reduction approaches}
\label{fig:heapdata}
\end{figure}

Figure~\ref{fig:heapdata} tabulates information that we have
gathered using the different implementations of head normalization
over the collection of examples illustrated in the previous
section. The two rows corresponding to each $\lambda$Prolog
program indicate, respectively, the number of internal term nodes
created and the number of bytes those terms occupied in the course
of executing the program; these figures are distinct because the
number of bytes needed for a given term node varies in the {\it
Teyjus} implementation depending on the type of the node. The
columns are to be understood as follows: {\it implicit
suspensions} corresponds to 
the reduction scheme where suspensions are recorded only in the
structure of recursive procedure calls, {\it explicit suspensions}
corresponds to the approach that explicitly realizes each rewrite
rule in Figure~\ref{fig:rewriterules} and the {\it combination
approach} represents the amalgamation of the other two.

The data in Figure~\ref{fig:heapdata} show that
the combination reduction approach
has significant superiority, especially in the cases of the first
four programs. Comparing the performance of the first reduction
strategy, which performs substitutions eagerly, with that of the
combination reduction strategy, which follows the same control regime
of the former one but performs the substitution lazily, the
advantage of the delayed substitution strategy is significant. In the
case of {\it compiler} and {\it hilbert}, the structure creation
using the combination reduction strategy is less than 5\% of that using
the eager substitution reduction strategy.

The better performance of the lazy substitution strategy is
attributable, ultimately, to the fact that delaying creates
substantially more opportunities for sharing in the structure
traversal required for substitution and reduction. As we discussed
in Section \ref{sec:imp-heap}, the eager substitution strategy has
potential drawbacks in the following two situations: first,
redices are generated dynamically across the computation steps,
which occurs frequently in the L$_\lambda$ programs; second,
redices are embedded in the term into which the substitutions have
to be performed. These drawbacks are avoided by the lazy
substitution strategies.

Towards understanding the enormous performance differences between
the first and the third reduction strategies in Figure
\ref{fig:heapdata}, we observe that structures having significant
quantities of embedded redices can be produced whenever the
programs embody an intrinsic use of higher-order unification. A
central part of this computation is that of positing substitutions
towards reconciling the differences between what are known as {\it
flexible-rigid} disagreement pairs, \ie a pair of terms of the
form
\begin{tabbing}
\qquad\=\kill
\>$\langle \lambdax{x_1} \ldots \lambdax{x_l} (F\app t_1\app \ldots \app t_n),
\lambdax{x_1} \ldots \lambdax{x_l} (c\app s_1\app \ldots \app s_m) \rangle$
\end{tabbing}
\noindent where $F$ is an instantiatable variable, $c$ is a
constant or an abstractable variable occurrence which is bound by
one of $x_1$,...,$x_l$ and $t_1,\ldots,t_n, s_1,\ldots,s_m$ are
arbitrary terms; we assume here that the binder lengths of the two
terms are identical, something that can be arranged based on
typing considerations under $\beta$-conversion.
Using the procedure due to Huet \cite{Huet75TCS}, a collection of
substitutions known as the imitation and projection substitutions
would be posited for $F$ in this situation. These substitutions
all have the structure
\begin{tabbing}
\qquad\=\kill
\>$\{ \langle F, \lambdax{w_1} \ldots \lambdax{w_n} (h \app (H_1 \app w_1 \app
\ldots \app w_n)\app \ldots \app (H_o\app w_1 \app \ldots \app w_n))\rangle
\}$
\end{tabbing}
\noindent where $h$ is either a constant or one of $w_1, \ldots,
w_n$ and $H_1, \ldots, H_o$ are new instantiatable variables. Now,
in subsequent steps of the computation, these new variables may
themselves become instantiated in a similar way yielding embedded
redices at all the places where $F$ appears. Moreover, the
instantiations for the variables $H_1, \ldots, H_o$ may themselves
contain embedded redices, resulting in further embedded redices in
the binding for $F$. Thus, after several invocations of the
unification computations, there are several embedded uncontracted
redices left in the binding determined for $F$ . An interesting
point to note is that these embedded redices all appear at the
argument position of the term serving as the binding of $F$. Thus,
each of these redices will be left in place by the head
normalization procedure whenever it is invoked to manifest the
top-level structure of a subterm in which the redices are
embedded. If the entire computation ends once a binding for $F$ is
determined, then the embedded redices are a harmless artifact and
do not have significantly influence performance characteristics.
However, in most cases, it is expected that the bindings found for
variables such as $F$ are used in further computations. In these
situations, the terms on which the substitutions are performed
have to be eventually (head) normalized. The performance
differences noted relative to our test suite owe significantly to
manifestations of this kind of phenomenon.

Theoretically, the eager substitution strategy has certain
benefits when the computation system involves backtracking.
However, in real executions, the benefits gained is not
significant enough to outweigh its other disadvantages we
discussed before. Backtracking is used to implement nondeterminism
and is realized in the following way. When there are multiple
branches the computation process can proceed to, the current
computation status and term structures under manipulation are
recorded, and then the computation process proceeds to one of the
possible branches; if failure is encountered on the chosen branch,
the computation procedure backtracks to the nearest choice point
by resetting the computation status and relevant term structures,
and then proceeds to the next possible branch. Since the first
reduction strategy tends to perform substitutions eagerly, it has
more opportunities to create new structures generated from the
substitutions on the heap before choice points. After
backtracking, these structures persist and certainly do not need
to be reset. If these term structures do not have redices embedded
inside, $\ie$ they are in their normal forms, they will not be
affected by the subsequent reductions, and their (head)
normalization will not create new terms on the heap. On the other
hand, if the lazy reduction strategies are used, the term
structures created before choice point may involve delayed
substitutions. Thus, each time after backtracking to this
choice point, the unreduced terms will be restored, and their
(head) normal forms will be rebuilt on the heap by later
reductions. However, if there {\it are} redices embedded inside
the structures on which substitutions will be performed, the terms
created by the eager substitution strategy are not normal forms
either, and later reductions will still build new terms. In this
situation, the eager substitution strategy gains no benefits, and
further the advantages of using lazy reduction strategies to
combine the substitution walks are significant. As we discussed
before, in most cases, there are complex redices embedded inside
the structures on which substitutions are performed. Thus, the
disadvantages of this eager substitution strategy we discussed
previously outweigh its benefit in a large degree. This is the
reason that although the two genuine unification programs require
frequent backtracking, their heap usage with the eager
substitution strategy is far worse than that with the combination
one.

The disadvantage of the eager substitution strategy is more obvious
when the structures of the arguments of the head normal forms are
complicated. As we discussed above, in general situations, the arguments
of a head normal form created by the eager substitution strategy
are overwritten by the following invocations of the head normalization
process, thereby being redundant.
It is apparent that the more complex these structures
are, the more heap cells are unnecessarily consumed by the head
normalization process using the eager substitution strategy.
This explains the difference of the improvements from the combination
strategy to the eager substitution one in different test cases
appears in Figure~\ref{fig:heapdata}. Specifically,
the ratios between the heap usage of the combination approach and
the implicit one are $19.66$ in the $compiler$ case, while this
number in the $typeinf$ case is $4.15$, although the
two programs both belong to the $L_{\lambda}$ subset.
A similar fact can be observed from the two programs of the genuine
higher-order class: the ratio is $30.15$ in the $hilbert$ case,
but only $1.15$ in $funtrans$. Such a difference of ratios is mainly
caused by the different sizes of the terms under manipulation of
these programs, which are reflected by the number of nodes of the
input terms. For instance, in the $compiler$ case, by controlling
the number of nodes of the input term, we obtain the heap usage
of the implicit and combination approaches shown in Figure~\ref{fig:progsize}.

\begin{figure}[!ht]
\begin{center}
\begin{tabular}{|l|l|l|l|l|} \hline
the number of nodes   &implicit suspensions   &combination approach
&ratio \\
in the input term     & & & \\\hline
31              &2,683       &625           &4.29 \\\hline
49              &7,670       &1,278         &6.00 \\\hline
103             &30,056      &3,222         &9.34 \\\hline
202             &101,204     &6,786         &14.91\\\hline
\end{tabular}
\end{center}
\caption{The effect of the term size}
\label{fig:progsize}
\end{figure}

\noindent
It can be observe from Figure~\ref{fig:progsize} that when the
size of the input term is small, the ratio is around $4$, which
is similar to that in the $typeinf$ case. However, when the
size of the input term increases, the disadvantage of the
eager substitution approach is exaggerated, and therefore
the ratio increases rapidly.

%% file: conclusion.tex
\chapter{Conclusions}\label{chp:conclusion}
We have examined different approaches to using explicit
substitutions in reduction computations in the thesis. The
computation contexts we are interested in are computation systems
of common metalanguages, logical frameworks and proof assistants.
In these situations, many substitutions may have to be performed
into the same subcomponent of a given object in the course of a
larger calculation, and these substitutions are effected by
creating and contracting at different points of time
$\beta$-redices that span over the relevant structure. We have
shown that the explicit substitution notation is useful in these
situations to combine such substitutions and avoid redundant
structure creation. In fact, most simplifiers for
$\lambda$-calculus actually take advantage of the explicit
substitution notation in spirit, as is manifest in their use of
environments and closures.

However, in the situations in which the term is known at the
beginning and is going to be eventually fully normalized, it is
beneficial to fully normalize the term using the explicit notation
only implicitly instead of normalizing the term in a demand-driven
manner as the head normalization strategies do, because heap space
used to record suspensions during the reduction process can be
avoided. \ignore{However, it is often believed that an intrinsic
use of explicit substitutions---in particular, a reflection of
substitution encodings into term structure and the creation of
such terms during reduction---can be costly in terms of space
usage. This argument may hold in the situations that the term is
known at the beginning and is going to be eventually fully
normalized. In this case, instead of normalizing the term in a
demand-driven manner as the head normalization strategies do, it
is beneficial to fully normalize the term using the explicit
notation only implicitly. As we argued above, in the computation
contexts we are interested in, the computation over
$\lambda$-terms often does not have such a character.}Further, if
the computation system involves non-deterministic search
situations implemented using backtracking, following the
discussion in Chapter \ref{chp:comparison}, the eagerness in
creating normal forms on the heap has potential benefits. The
benefits of the eager substitution strategy are not obvious
because in most cases, the structures on which substitutions are
performed have redices embedded inside. Taking into account this
factor together with the possible benefits of using explicit
substitution only implicitly, it is natural to consider a
reduction strategy which reduces the term eagerly. Following the
control regime of the eager substitution strategy, at the point
where the head of a head normal form is exposed and there are
delayed substitutions to be performed on the arguments of that
head normal form, instead of actually carrying out the
substitutions or explicitly building suspensions over those
arguments, the reduction procedure recursively calls itself to
normalize the arguments under the delayed substitutions. In this
sense, the (sub)term is fully normalized if there are suspended
substitutions involving it. It is obvious that this reduction
strategy has more opportunities to create normal forms before
choice points than the eager substitution and lazy reduction ones.
However, if the subsequent computation on the term do not require
its normal form to be fully exposed, for example, the failure of
comparison between two unequal terms can be detected before their
normal forms are fully revealed, the effort spent on normalizing
the unneeded parts of the terms is redundant. This is in fact a
tradeoff: if a term remains fixed till a successful path is found
or until its full normal form is needed, it is beneficial to
reduce it completely before a backtracking point is encountered so
that rollbacks in computation do not cause such reductions to be
undone and subsequently redone; on the other hand, if only part of
the term is needed possibly because its structure is never fully
examined in a successful computation or because failure occurs
after only part of it is examined, the lazy substitution
strategies have a better chance to avoid traversing the term
completely and normalizing its unneeded parts. While a detailed
analysis of this problem is beyond this thesis, we mention that
studies have been conducted subsequent to the work in this thesis
towards the precise manner in which these two factors impact
behavior in practice~\cite{LNX03}. The experiment results show
that the heap usage of the eager reduction strategy is comparable
to that of the combined one, and in some test cases is even
slightly better, which implies that in real programs the first
situation occurs more frequently.

To further improve the heap usage of our reduction strategies, we
observe the fact that if a term $t$ is closed, it will not be
affected by the substitutions generated from the reductions of the
structures enclosing it. In particular, in this case, a suspension of 
form $\env{t,ol,nl,env}$ can be simplified to $t$. If we can recognize
such a closed term before traversing it, then the effort spent on
the traversal over this term for substitution performance can be
saved. In fact, there is a variation of the suspension notation
which associates annotations with terms to denote their
closedness~\cite{nadathur99finegrained}. Annotations have
different impacts on the heap usage of the different reduction
strategies according to their ability to combine substitutions. We
conducted experiments comparing the heap usage of these reduction
strategies with annotations and the experimental data show that
with annotations, the heap usage of the eager substitution
strategy is improved significantly in most test cases, while the
improvements of the lazy substitution and lazy reduction
strategies are relatively less. The reason for this difference is
the following. As we discussed in Chapter \ref{chp:comparison},
the structures produced by higher-order unification often have
significant quantities of embedded redices. Such a redex is
generated from the binding of a free variable, say $F$, in an
environment such as $\lambda{x_1}...\lambda{x_l}(...(F\app t_1\app
... \app t_n)...)$, to a term of form $\lambdax{w_1} \ldots
\lambdax{w_m}\app s$. It can be observed that the redex formed
after this substitution is closed, $\ie$, it would not be affected
by the delayed substitutions over it. Thus annotations can help
the eager substitution strategy avoid the redundant term
traversals purely for substitution performance, over such redices.
Note that the reduction walk over such term structures cannot be
avoided, since there are redices embedded inside them. However,
the lazy substitution strategies avoid separate substitution and
reduction walks by delaying those substitutions to be performed
along with the reduction steps, while the eager reduction strategy
avoids such separate walks by eagerly performing reductions along
with the substitution traversals. Since these redices have to
eventually be normalized, the improvements gained by adding
annotations to these reduction strategies are not as much as that
of the eager substitution one. While a detailed analysis of this
problem is beyond this thesis, we conducted the subsequent studies 
towards the precise manner in which the
annotations impact behavior in practice in~\cite{LNX03}.

Our focus in this work has been mainly on a comparison of space
usage and the elimination of redundant structure creation. Another
important factor to consider is the time efficiency of each of the
reduction approaches. The procedure based on the naive view of
rewriting is the simplest to realize and the {\it Teyjus} system,
in fact, embodies an iterative rendition of this procedure using a
term stack. Adapting such an optimized implementation to the other
reduction strategies, and including a way in which garbage
collection costs are taken into account, it is meaningful to
obtain and compare the time measurements of these reduction
strategies. A different aspect that is relevant to study concerns
the compiled realization of reduction. Recent work relative to the
{\it Coq} system has shown how to use compilation assuming eager
reduction and substitution strategies to obtain substantial
speedups in comparison with the existing interpretive approach
\cite{GL02icfp}. The examples considered in this study seem to be
ones where the terms to be normalized are available in complete
form at the beginning of the computation. As we have argued, this
situation is different from what is encountered in metalanguages
such as $\lambda$Prolog and Elf. It is, therefore, of interest to
see if explicit substitutions can be built into a compilation
model towards harnessing the benefits of laziness in substitution
over and above those of compilation in these contexts.

%% file: master_CoRR.bbl
\begin{thebibliography}{10}

\bibitem{ACCL91}
Mart{\'{\i}}n Abadi, Luca Cardelli, Pierre-Louis Curien, and Jean-Jacques
  L{\'{e}}vy.
\newblock Explicit substitutions.
\newblock {\em Journal of Functional Programming}, 1(4):375--416, 1991.

\bibitem{Bar81}
H.~P. Barendregt.
\newblock {\em The Lambda Calculus: Its Syntax and Semantics}.
\newblock North Holland Publishing Co., 1981.

\bibitem{BBLR96}
Z.~Benaissa, D.~Briaud, P.~Lescanne, and J.~Rouyer-Degli.
\newblock $\lambda\upsilon$, a calculus of explicit substitutions which
  preserves strong normalization.
\newblock {\em Journal of Functional Programming}, 6(5):699--722, 1996.

\bibitem{debruijn72}
{N. de} Bruijn.
\newblock Lambda calculus notation with nameless dummies, a tool for automatic
  formula manipulation, with application to the {Church-Rosser Theorem}.
\newblock {\em Indag. Math.}, 34(5):381--392, 1972.

\bibitem{deBruijn80}
{N. de} Bruijn.
\newblock A survey of the project {AUTOMATH}.
\newblock In J.~P. Seldin and J.~R. Hindley, editors, {\em To {H. B. Curry}:
  Essays on Combinatory Logic, Lambda Calculus and Formalism}, pages 579--606.
  Academic Press, 1980.

\bibitem{Cons86}
R.~L. Constable, S.~F. Allen, H.~M. Bromley, W.~R. Cleaveland, J.~F. Cremer,
  R.~W. Harper, D.~J. Howe, T.~B. Knoblock, N.~P. Mendler, P.~Panangaden, J.~T.
  Sasaki, and S.~F. Smith.
\newblock {\em Implementing Mathematics with the Nuprl Proof Development
  System}.
\newblock Prentice-Hall, 1986.

\bibitem{CH88}
Thierry Coquand and G\'erard Huet.
\newblock The calculus of constructions.
\newblock {\em Information and Computation}, 76(2/3):95--120, February/March
  1988.

\bibitem{Dowek93tr}
Gilles Dowek, Amy Felty, Hugo Herbelin, G{\'e}rard Huet, Chet Murthy, Catherine
  Parent, Christine Paulin-Mohring, and Benjamin Werner.
\newblock The {Coq} proof assistant user's guide.
\newblock Rapport Techniques 154, INRIA, Rocquencourt, France, 1993.
\newblock Version 5.8.

\bibitem{GL02icfp}
B.~Gr\'{e}goire and X.~Leroy.
\newblock A compiled implementation of strong reduction.
\newblock In {\em Proceedings of the Seventh ACM SIGPLAN International
  Conference on Functional Programming}, pages 235--246, Pittsburgh, October
  2002.

\bibitem{Harper86ECS-LFCS-86-14}
R.~Harper.
\newblock Introduction to {S}tandard {ML}.
\newblock Technical Report ECS-LFCS-86-14, Laboratory for Foundations of
  Computer Science, University of Edinburgh, November 1986.
\newblock Revised by {N}ick {R}othwell, January 1989, with exercises by Kevin
  Mitchell.

\bibitem{HHP93}
Robert Harper, Furio Honsell, and Gordon Plotkin.
\newblock A framework for defining logics.
\newblock {\em Journal of the ACM}, 40(1):143--184, 1993.

\bibitem{Hindley86ILC}
J.~Roger Hindley and Jonathan~P. Seldin.
\newblock {\em Introduction to Combinators and Lambda-Calculus}.
\newblock Cambridge University Press, 1986.

\bibitem{Huet75TCS}
G.~Huet.
\newblock A unification algorithm for typed $\lambda$-calculus.
\newblock {\em Theoretical Computer Science}, 1:27--57, 1975.

\bibitem{KR97}
Fairouz Kamareddine and Alejandro R\'{i}os.
\newblock Extending the $\lambda$-calculus with explicit substitution which
  preserves strong normalization into a confluent calculus on open terms.
\newblock {\em Journal of Functional Programming}, 7(4):395--420, 1997.

\bibitem{LNX03}
Chuck Liang, Gopalan Nadathur, and Xiaochu Qi.
\newblock Choices in representation and reduction strategies for lambda terms
  in intensional context.
\newblock Technical Report 2003/02, Department of Computer and Information
  Science, University of Minnesota, October 2003.

\bibitem{Miller91jlc}
Dale Miller.
\newblock A logic programming language with lambda-abstraction, function
  variables, and simple unification.
\newblock {\em Journal of Logic and Computation}, 1(4):497--536, 1991.

\bibitem{nadathur99finegrained}
Gopalan Nadathur.
\newblock A fine-grained notation for lambda terms and its use in intensional
  operations.
\newblock {\em Journal of Functional and Logic Programming}, 1999(2), 1999.

\bibitem{NM88}
Gopalan Nadathur and Dale Miller.
\newblock An overview of {$\lambda$Prolog}.
\newblock In Kenneth~A. Bowen and Robert~A. Kowalski, editors, {\em Fifth
  International Logic Programming Conference}, pages 810--827. MIT Press,
  August 1988.

\bibitem{NM99cade}
Gopalan Nadathur and Dustin~J. Mitchell.
\newblock System description: Teyjus---a compiler and abstract machine based
  implementation of $\lambda${P}rolog.
\newblock In Harald Ganzinger, editor, {\em Automated Deduction--{CADE}-16},
  number 1632 in Lecture Notes in Artificial Intelligence, pages 287--291.
  Springer-Verlag, July 1999.

\bibitem{NX03}
Gopalan Nadathur and Xiaochu Qi.
\newblock Explicit substitutions in the reduction of lambda terms.
\newblock In {\em Proceedings of the 5th ACM SIGPLAN international conference
  on Principles and practice of declaritive programming}, pages 195--206. ACM
  Press, 2003.

\bibitem{NW98tcs}
Gopalan Nadathur and Debra~Sue Wilson.
\newblock A notation for lambda terms: A generalization of environments.
\newblock {\em Theoretical Computer Science}, 198(1-2):49--98, 1998.

\bibitem{Paulson94}
Lawrence~C. Paulson.
\newblock {\em Isabelle: A Generic Theorem Prover}, volume 828 of {\em Lecture
  Notes in Computer Science}.
\newblock Springer Verlag, 1994.

\bibitem{Pfenning94cade}
Frank Pfenning.
\newblock Elf: A meta-language for deductive systems.
\newblock In A.~Bundy, editor, {\em Proceedings of the 12th International
  Conference on Automated Deduction}, pages 811--815, Nancy, France, June 1994.
  Springer-Verlag LNAI 814.
\newblock System abstract.

\bibitem{shao98:imp}
Z.~Shao, C.~League, and S.~Monnier.
\newblock Implementing typed intermediate languages.
\newblock In {\em Proc. 1998 {ACM} {SIGPLAN} International Conference on
  Functional Programming ({ICFP}'98)}, pages 313--323. ACM Press, September
  1998.

\end{thebibliography}
